\documentclass[useAMS,usenatbib]{mn2e}

\usepackage{graphicx}
\usepackage{times}
\usepackage{natbib}
\usepackage{rotating}
\usepackage{lastpage}

\newcommand{\be}{\begin{equation}}
\newcommand{\ee}{\end{equation}}
\newcommand{\ba}{\begin{eqnarray}}
\newcommand{\ea}{\end{eqnarray}}
\newcommand{\brr}{\begin{array}}
\newcommand{\err}{\end{array}}
\newcommand{\bc}{\begin{center}}
\newcommand{\ec}{\end{center}}

\newcommand{\mincir}{\raise
  -2.truept\hbox{\rlap{\hbox{$\sim$}}\raise5.truept \hbox{$<$}\ }}
\newcommand{\magcir}{\raise
  -2.truept\hbox{\rlap{\hbox{$\sim$}}\raise5.truept \hbox{$>$}\ }}
\newcommand{\siml}{\raise
  -2.truept\hbox{\rlap{\hbox{$\sim$}}\raise5.truept \hbox{$<$}\ }}
\newcommand{\simg}{\raise
  -2.truept\hbox{\rlap{\hbox{$\sim$}}\raise5.truept \hbox{$>$}\ }}

\title[Constraints on FRBs from cosmological simulations]{
Constraints on the distribution and energetics of fast radio bursts using cosmological hydrodynamic simulations}

\author[K. Dolag et al.]
{  K.~Dolag,$^{1,2}$\thanks{E-mail: dolag@usm.uni-muenchen.de}
B.~M. Gaensler,$^{3,4,5}$ A.~M. Beck$^{1,2}$ and M.~C. Beck$^6$\\
$^1$ University Observatory Munich, Scheinerstr. 1, 81679 Munich, Germany\\
$^2$ Max-Planck-Institut f\"ur Astrophysik, Karl-Schwarzschild Strasse
  1, 85748 Garching bei M\"unchen, Germany\\
$^3$ Sydney Institute for Astronomy, School of Physics, The University of Sydney, NSW 2006, Australia\\
$^4$ ARC Centre of Excellence for All-sky Astrophysics (CAASTRO) \\
$^5$ Dunlap Institute for Astronomy and Astrophysics, The University of Toronto, Toronto ON M5S~3H4, Canada\\
$^6$ Department of Physics, University of Konstanz, Universit\"{a}tsstr. 10, D-78457 Konstanz, Germany\\
}

\begin{document}

\date{Accepted ???. Received ???; in original form ???}

\pagerange{\pageref{firstpage}--\pageref{LastPage}} \pubyear{0000}

\maketitle

\label{firstpage}

\begin{abstract}
We present constraints on the origins of fast radio bursts (FRBs) using
large cosmological simulations.  We calculate contributions to FRB
dispersion measures (DMs) from the Milky Way, from the local Universe, from
cosmological large-scale structure, and from potential FRB host galaxies,
and then compare these simulations to the DMs of observed FRBs. We find that
the Milky Way contribution has previously been underestimated by a factor of
$\sim2$, and that the foreground-subtracted DMs are consistent with a
cosmological origin, corresponding to a source population observable to a
maximum redshift $z\sim0.6-0.9$. We consider models for the spatial
distribution of FRBs in which they are randomly distributed in the Universe,
track the star-formation rate of their host galaxies, track total stellar
mass, or require a central supermassive black hole. Current data do not
discriminate between these possibilities, but the predicted DM distributions
for different models will differ considerably once we begin detecting FRBs
at higher DMs and higher redshifts. We additionally consider the
distribution of FRB fluences, and show that the observations are consistent
with FRBs being standard candles, each burst producing the same radiated
isotropic energy.  The data imply a constant isotropic burst energy of
$\sim$$7\times10^{40}$~erg if FRBs are embedded in host galaxies, or
$\sim$$9\times10^{40}$~erg if FRBs are randomly distributed. These energies
are 10--100 times larger than had previously been inferred. Within the
constraints of the available small sample of data, our analysis favours FRB
mechanisms for which the isotropic radiated energy has a narrow distribution
in excess of $10^{40}$~erg.
\end{abstract}



\section{Introduction} 
\label{sec_intro}

Fast radio bursts (FRBs) are a newly identified and as-yet-unexplained class
of transient objects
\citep{2007Sci...318..777L,2013Sci...341...53T,2014ApJ...797...70K}. The ten
known FRBs currently in the literature are characterised by short
($\approx1$~ms), bright ($\ga$1~Jansky) bursts of radio emission;
none have been seen to repeat, and all but two occurred at high Galactic
latitude, $|b| > 20^\circ$. The implied all-sky event rate is enormous,
around 10\,000 per day \citep{2013Sci...341...53T}.

The radio signals from FRBs experience a frequency-dependent dispersion
delay as they propagate through ionised gas, just as is routinely seen for
radio pulsars. However, for most observed FRBs, the very high dispersion
measures (DMs), in the range 400--1100~pc~cm$^{-3}$, are more than an order
of magnitude larger than the DM contribution expected from the interstellar
medium (ISM) of the Milky Way in these directions. The currently favoured
interpretation is that the observed DMs seen for FRBs correspond primarily
to free electrons in the intergalactic medium (IGM) along the line of sight
\citep{2014ApJ...785L..26L,2014MNRAS.443L..11D},
with an additional but presumed small contribution from any host galaxy
\citep{2013Sci...341...53T}.  Simple assumptions about the density of the IGM then immediately
imply that FRBs are at cosmological distances, corresponding to redshifts in
the range 0.5--1 \citep{2013Sci...341...53T}.

The nature of FRBs is not known. Possibilities that have been proposed
include flaring magnetars \citep{Popov2007,Popov2013,2014MNRAS.442L...9L},
mergers of binary neutron stars \citep{Totani2013},
gamma-ray bursts \citep{Zhang2014}, collisions between neutron stars and
asteroids/comets \citep{2015arXiv150205171G}, or the collapse of supramassive neutron stars
\citep[``blitzars'';][]{Falcke2014}.
There are two approaches through which we can
make further progress in discerning between these and other possibilities,
One approach is to localise individual FRBs, so that we can then identify
multi-wavelength counterparts, host galaxies, afterglows and redshifts.
However, such data are not yet available, because all FRBs seen so far have
been poorly localised, and most were not detected in the data until months
or years after they were observed. The  alternative is to consider the
ensemble properties of FRBs, and to compare
these to different simulated FRB distributions. 
In this paper we adopt the latter approach, in which we use state-of-the-art
hydrodynamic simulations to consider synthesised populations of FRBs
within a cosmological volume. We consider a series of simple
assumptions as to the way in which FRBs are distributed relative to the 
distribution of large-scale structure, compute corresponding distributions
of DM and fluence, and compare these to the observations. In
\S\ref{sec_obs} we summarise the observed properties of the ten published
FRBs. In
\S\ref{sec_signal} we consider the various foreground contributions to the
observed FRB DMs,
including the Milky Way's disk and spiral arms (\S\ref{sec_ne2001}), the Galactic halo
(\S\ref{sec_halo}) and the local Universe (\S\ref{sec_lu}).
In \S\ref{sec_cosmo} we then calculate the expected cosmological component of FRB DMs
using the {\em Magneticum Pathfinder}\ simulation, and compare this to
observations. 
In \S\ref{sec_fluence} we compare simulated and observed 
FRB fluences 
in order to constrain the isotropic energy released in the radio bursts.


\section{Observations of Fast Radio Bursts} 
\label{sec_obs}

The observational data we consider are the ten published FRBs as listed in
Table~\ref{tab:obs},
for each of which we provide Galactic coordinates, the
observed value of DM, peak flux and fluence, and the central observing
frequency at which the FRB was detected.

\begin{table*}
\caption{Observed and derived properties of the ten published FRBs. The
first seven columns list the observed parameters of each burst, as drawn
from the references given in the eighth column. Note that
\citet{2015MNRAS.447.2852K} provide revised peak fluxes and fluences
for nine of the ten published FRBs, and further note that these revised
values are further underestimates typically by a factor of $\sim2$. In this
Table, we have used the revised values of \citet{2015MNRAS.447.2852K} where
available, and have further doubled all values of fluxes and fluences.  DM$_{\rm ISM}$, DM$_{\rm
halo}$ and DM$_{\rm local~sim}$ list the inferred contributions to the total
DM from the Galactic disk and spiral arms (\S\ref{sec_ne2001}), Galactic halo (\S\ref{sec_halo}) 
and local Universe (\S\ref{sec_lu}), respectively. In the
final column, DM$_{\rm cosmo}$ lists the dispersion measure contribution remaining after
subtracting DM$_{\rm ISM}$ and  DM$_{\rm halo}$ (but not DM$_{\rm local~sim}$) from the observed value.
As discussed in \S\ref{sec_lu}, we assume DM$_{\rm LU} = 0$ throughout.}

\label{tab:obs}
\begin{tabular}{|l c c c c c c c|||| c c c c|} 
\hline
FRB & $\ell$ & $b$ & DM$_\mathrm{obs}$ & Peak flux & Fluence & Freq. & Ref. & DM$_{\rm ISM}$ & DM$_{\rm halo}$ & DM$_{\rm local~sim}$ &
DM$_\mathrm{cosmo}$  \\
   & ($^\circ$) & ($^\circ$)& (pc~cm$^{-3}$) & (Jy) & (Jy~ms) & (GHz) & & (pc~cm$^{-3}$) & (pc~cm$^{-3}$)
& (pc~cm$^{-3}$) & (pc~cm$^{-3}$)  \\
\hline
010125$^1$ & 356.6 & --20.0 & $790   \pm3$   & 1.10$^{+0.22}_{-0.16}$ & 11.2$^{+6.0}_{-4.0}$ & 1.4 & 1,2 & 110 & 30 & 13  & 650 \\
010621 & 25.4  & --4.0  & $746\pm1$          & 1.04$^{+0.26}_{-0.22}$ & 8.6$^{+7.2}_{-3.8}$  & 1.4 & 3,2 & 537 & ---& --- & --- \\
010724 & 300.8 & --41.9 &  $375 \pm1$        & $>3.16$             & $>63.0$           & 1.4 & 4,2 & 44  & 30 & 20  & 301 \\
110220 & 50.8  & --54.7 & $944.38\pm0.05$    & 2.22$^{+2.24}_{-0.20}$  & 14.6$^{+4.8}_{-3.4}$ & 1.3 & 5,2 & 35  & 30 & 5   & 879  \\
110626 & 355.8 & --41.7 & $723.0 \pm0.3$     & 1.26$^{0.40}_{-0.26}$  & 1.8$^{+2.6}_{-0.4}$  & 1.3 & 5,2 & 47  & 30 & 10  & 646 \\
110703 & 81.0  & --59.0 &$1103.6 \pm0.7$     & 0.9$^{+0.42}_{-0.20}$  & 3.6$^{+4.6}_{-2.2}$  & 1.3 & 5,2 & 33  & 30 & 14  & 1041 \\
120127 & 49.2  & --66.2 & $553.3 \pm0.3$     & 1.24$^{+0.26}_{-0.20}$ & 1.6$^{+1.2}_{-0.6}$  & 1.3 & 5,2 & 32  & 30 & 9   & 491 \\
121102 & 175.0 & --0.2  & $557 \pm2$         & 0.8$^{+0.8}_{-0.2}$         & 2.4$^{+8.0}_{-2.0}$      & 1.4 & 6,2 & 192 & 30 & 10  & 335 \\
131104 & 260.6 & --21.9 &$778.5^{+0.2}_{-0.3}$  & 2.2$^{+0.1}_{-0.2}$       & 1.9$^{+0.08}_{-0.18}$ & 1.4 & 7   & 71 & 30 & 10  & 678   \\
140514 & 50.8  & --54.6 & $562.7 \pm0.6$     & 0.94$^{+0.22}_{-0.16}$ & 2.6$^{+4.6}_{-1.0}$   & 1.4 & 8,2 & 35  & 30 & 5   & 498  \\
\hline
\end{tabular}
References: 
(1)~\citet{2014arXiv1407.0400B};
(2)~\citet{2015MNRAS.447.2852K};
(3)~\citet{2012MNRAS.425L..71K};
(4)~\citet{2007Sci...318..777L}; 
(5)~\citet{2013Sci...341...53T};
(6)~\citet{2014arXiv1404.2934S};
(7)~\citet{2014arXiv1412.1599R};
(8)~\citet{2014arXiv1412.0342P}.

$^1$This burst was misnamed FRB~011025 by \citet{2014arXiv1407.0400B}.

\end{table*} 

\section{Foreground Dispersion Measure}
\label{sec_signal}

\begin{figure}
\includegraphics[width=0.425\textwidth]{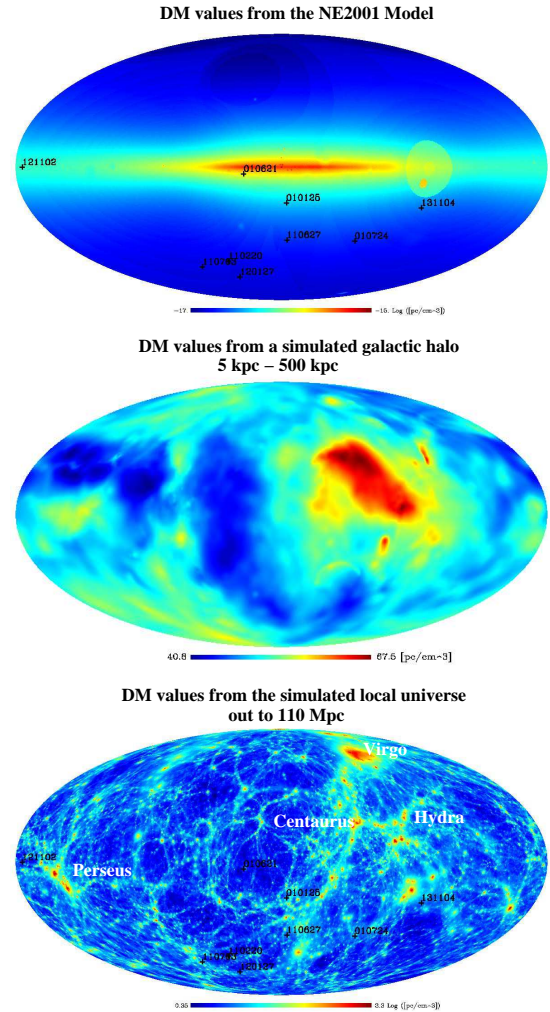}
\caption{Full-sky maps of the DM contributions of various foregrounds, in
Galactic coordinates centred on $\ell = 0$, $b = 0$.
From top to bottom: DM$_{\rm ISM}$ predicted by the NE2001 model of
\citet{2002astro.ph..7156C,2003astro.ph..1598C}; DM$_{\rm halo}$ contribution
from a typical simulation of the extended halo of a Milky-Way-type galaxy;
DM$_{\rm LU}$ contribution from the simulated local Universe out to a distance of 110~Mpc,
showing prominent local large-scale structures such as the Perseus-Pisces region,
the Virgo cluster and the Centaurus supercluster region. 
The positions of the FRBs in Table~\ref{tab:obs} are indicated in the top and
bottom panels (FRB~140514 is at almost the same position as FRB~110220 and
is not shown.)
}
\label{fig:foreground}
\end{figure}

The DMs observed for FRBs as listed in Table~\ref{tab:obs} must consist of several contributions,
corresponding to the various 
astrophysical structures through which the radio signal has traversed. For
our purposes, the DM contributions both from the Milky Way Galaxy and 
from local large-scale structure 
are considered to be foregrounds, 
which we would like to remove to isolate the cosmological signal. Some of
the foreground components  are difficult to obtain from 
direct observations, and we use cosmological simulations
to estimate the contribution of these components. We define the observed dispersion
measure, DM$_\mathrm{obs}$, as:
\begin{equation}
{\rm DM}_\mathrm{obs} = {\rm DM}_\mathrm{ISM} + {\rm DM}_\mathrm{halo} + 
{\rm DM}_\mathrm{LU} + 
{\rm DM}_\mathrm{LSS} + {\rm DM}_\mathrm{host},
\label{eqn_dmsum}
\end{equation}
where DM$_{\rm ISM}$  is the contribution from the Milky Way disk and
spiral arms, DM$_{\rm halo}$ is that from the Galactic halo,
DM$_{\rm LU}$ is that from the local Universe,
DM$_{\rm LSS}$ is that from large-scale structure and
DM$_{\rm host}$ is any contribution from any host galaxy or other immediate
environment of the FRB. We define:
\begin{equation}
{\rm DM}_{\rm cosmo} = {\rm DM}_\mathrm{LSS} + {\rm DM}_\mathrm{host}
\label{eqn_cosmo}
\end{equation}
as the signal to be estimated from cosmological hydrodynamic simulations,
with the remaining terms on the right-hand side of
Equation~(\ref{eqn_dmsum}) representing the foreground signal that must be
accounted for in order to derive ${\rm DM}_{\rm cosmo}$ from DM$_{\rm obs}$.
Fig.~\ref{fig:foreground} shows full sky-maps
of the DM contributions of the various foreground components, as discussed
in more detail in \S\S\ref{sec_ne2001}, \ref{sec_halo} and \ref{sec_lu}.

\subsection{The Galaxy Model}
\label{sec_ne2001}

We calculate the foreground contribution from the Milky Way's disk and spiral arms using the widely-used NE2001 distribution
\citep{2002astro.ph..7156C,2003astro.ph..1598C}. This three-dimensional
model of thermal electron density, $n_{e,{\rm NE2001}}$, is based on the DMs observed for
Galactic radio pulsars,  and includes contributions from axisymmetric thin
and thick disks,  Galactic spiral arms, and small-scale elements such as
local under-dense regions and localised high density clumps. 

We implement the NE2001 model using the {\sc hammurabi}\ code
\citep{2009A&A...495..697W} to integrate the Galactic electron
density along a given
sightline from the position of the Sun out to the limit of the model:
\begin{equation}
\mathrm{DM}_\mathrm{ISM}=\int_0^\infty n_{e,\mathrm{NE2001}} \,\mathrm{d}l~.
\end{equation}
In Fig.~\ref{fig:foreground} we show the full-sky DM signal for the NE2001
model, together with the positions of the FRBs in Table~\ref{tab:obs}.  We
list our values derived for DM$_{\rm ISM}$ in Table~\ref{tab:obs} --- 
the observed DMs far exceed DM$_{\rm ISM}$ except in the case of
FRB~010621, where the two values are comparable. For this reason FRB~010621
has been argued to be of possible Galactic origin
\citep{2014MNRAS.440..353B}, and we exclude it from further consideration.

The estimates that we have derived for DM$_{\rm ISM}$ match the
corresponding values given in the referenced papers, with the exception of
FRB~010724 for which \citet{2007Sci...318..777L} quoted  DM$_{\rm ISM} =
25$~pc~cm$^{-3}$ compared to DM$_{\rm ISM} = 44$~pc~cm$^{-3}$ as given by
NE2001 and listed here.  We note that the NE2001 model is known not to give
reliable estimates for pulsar distances or DMs at high Galactic latitudes,
$|b| \ga 40^\circ$, as discussed extensively by \citet{2008PASA...25..184G}.
However, this has minimal impact on our estimates of DM$_{\rm ISM}$, which
is integrated to the edge of the distribution. At high latitudes, the
results of NE2001 can be roughly approximated as DM$_{\rm ISM} \sim 30 /
|\sin b|$~pc~cm$^{-3}$, while \citet{2008PASA...25..184G} adopts DM$_{\rm
ISM} = 26 / |\sin b|$~pc~cm$^{-3}$. The difference between these two options
is typically $\la1\%$ of DM$_{\rm obs}$ for FRBs.

\subsection{The Halo Model}
\label{sec_halo}

The values of  DM$_{\rm ISM}$ calculated in \S\ref{sec_ne2001} above only
account for the foreground DM originating in the disk structure of the Milky
Way; the NE2001 model lacks the contribution of a virialised dark
matter halo with a hot gaseous atmosphere.  As per
Equation~(\ref{eqn_dmsum}), we must also consider the free-electron
contribution of the surrounding Galactic halo, which we model using
numerical simulations.  To estimate DM$_{\rm halo}$, we use a cosmological
simulation of a representative Milky Way-type galactic halo including hot
thermal electrons.  We use the existing simulations of
\cite{2013MNRAS.435.3575B} to estimate the contribution to the DM.
Fig.~\ref{fig:foreground} shows an all-sky projection of the corresponding
DM distribution.

The simulation of \cite{2013MNRAS.435.3575B} is based on initial
conditions which were originally introduced by
\cite{2002MNRAS.335L..84S}.  Briefly, the simulation is based on a
large cosmological box with initial fluctuations of power spectrum index
$n = 1$ and a fluctuation amplitude $\sigma_{8}
= 0.9$, in which a Milky Way-like dark matter halo is identified.  We
use the fully magneto-hydrodynamic simulation labeled GA2, which
contains 1,055,083 dark matter particles inside the virial radius at
present redshift.  The halo is comparable in mass ($\approx
3\times10^{12}$~M$_\odot$) and in virial size ($\approx 270$~kpc) to the
halo of the Milky Way.  It does not undergo any major mergers after a
redshift $z\approx2$ and also hosts a sub-halo population
comparable to the satellite population of our own Galaxy.
Additionally, we follow the gas and stellar components by including
multi-phase gas particles, which follow the prescriptions of radiative
cooling, supernova feedback and star formation based on the work of
\cite{2003MNRAS.339..289S} but without galactic winds.  Furthermore,
the simulation is extended with magnetic fields following the
implementation of \cite{2009MNRAS.398.1678D}.  In addition,
\cite{2013MNRAS.435.3575B} further extend the original magnetohydrodynamic
calculation of \citet{2012MNRAS.422.2152B} with a numerical
sub-grid model for the self-consistent seeding of magnetic fields by
supernova explosions.

\begin{figure*}
\includegraphics[width=0.33\textwidth]{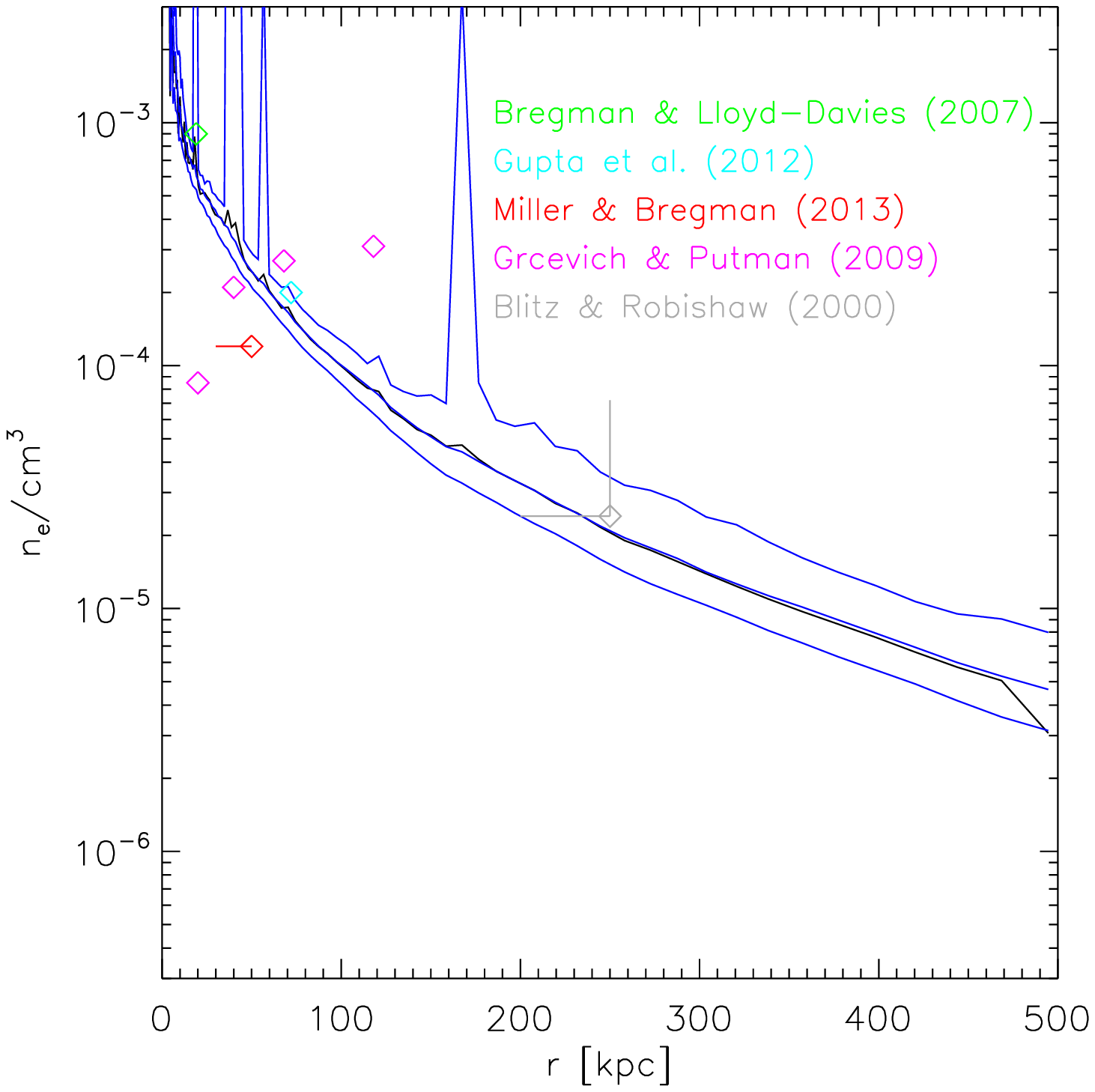}
\includegraphics[width=0.33\textwidth]{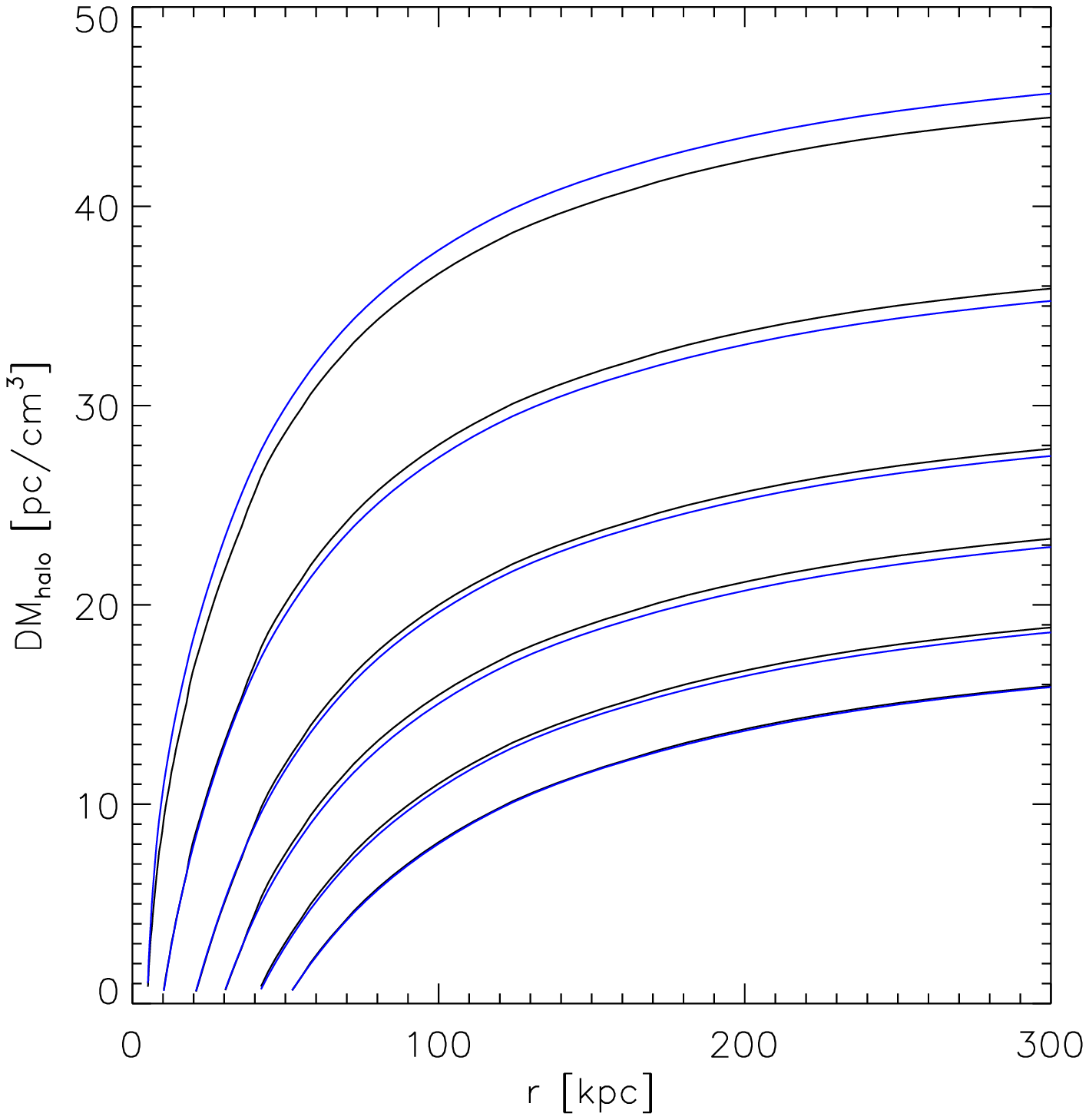}
\includegraphics[width=0.33\textwidth]{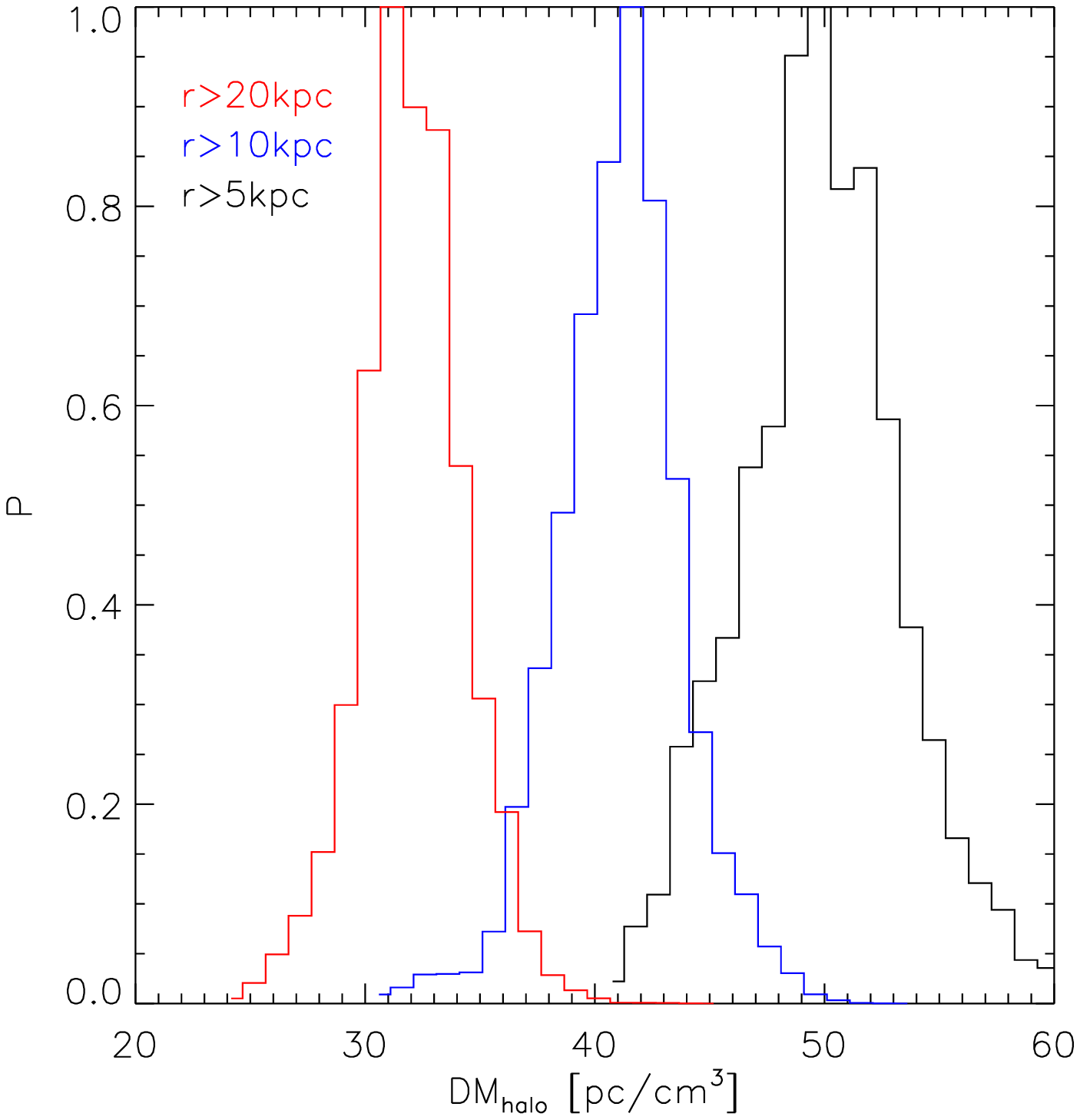}
\caption{Predicted DM properties of a hot halo in a cosmological
simulation of a Milky-Way-type galaxy.  The left panel shows calculations
for the radial distribution of free electrons, plus a comparison with
observational constraints for the Milky Way.  The black line shows the
electron density  obtained using the mean gas mass within each cell under
the assumption of full ionisation of a primordial gas composition, while the
blue lines show the median values, enveloped by the 10th and 90th
percentiles of the electron density distribution within the simulation.  The
spikes in the simulated density profiles are caused by individual
sub-structures within the simulated halo. The middle panel shows the
corresponding integrated DM, the different lines indicating  different inner
radii used as the starting point for the integration. The right panel shows
the predicted distribution of DM$_{\rm halo}$ over all possible sightlines, for three
possible values of the inner radius.
}
\label{fig:halo}
\end{figure*}

The left panel of Fig.~\ref{fig:halo} shows some resulting
predictions for
radial electron
density distributions representative of the Galactic halo, overplotted with observational data and
constraints covering the entire virial radius, as presented by
\cite{2013ApJ...770..118M}, \cite{2009ApJ...696..385G},
\cite{2000ApJ...541..675B}, \cite{2007ApJ...669..990B} and
\cite{2012ApJ...756L...8G}.
Our simulated radial electron number density profile also agrees closely
with the recent work of \cite{2014MNRAS.441.2593N}, which is based
on a constrained simulation of the Local Group. 

The middle panel of Fig.~\ref{fig:halo} shows the cumulative increase of
DM$_{\rm halo}$ as function of distance for different starting radii, while the right
panel shows the distribution of DM$_{\rm halo}$ over all sightlines for these radii.
At small radii, our results are consistent with the constraints from pulsar
DMs discussed in \S4.6 of \citet{2008PASA...25..184G}. Most interestingly,
we find that the electrons in the hot Galactic halo make a non-negligible
contribution to the total observed DMs for FRBs.

Our simulation of the halo has not been constrained to match the particular
structure of the Milky Way, so we cannot calculate specific values of
DM$_{\rm halo}$ for individual FRB sightlines as we did for DM$_{\rm
ISM}$ in \S\ref{sec_ne2001}. Rather, we use the results of Fig.~\ref{fig:halo} to determine an
indicative value for DM$_{\rm halo}$. To make such an estimate, 
we should ideally use a radius 
to begin the integration
that corresponds to the
outer edge of the NE2001 model. However, this outer edge is direction-dependent and
is not a well-defined concept. Given that the maximal extent of NE2001
from the Galactic centre is $\sim17$~kpc \citep[see Fig.~3 and Table~3
of][]{2002astro.ph..7156C}, we adopt from Fig.~\ref{fig:halo}
a representative halo electron column DM$_{\rm halo} =
30$~pc~cm$^{-3}$, as listed for all relevant FRBs in
Table~\ref{tab:obs}. 

\subsection{The Local Universe}
\label{sec_lu}

To consider the possible contribution of local superclusters to the DMs
observed for
FRBs, we use the final output of a cosmological hydrodynamic
simulation of the local Universe. Our initial conditions are similar to those
adopted by \citet{Mathis:2002} in their study
(based on a pure N-body simulation) of structure formation in the local
Universe.  We first apply a gaussian smoothing to the galaxy distribution in the IRAS 1.2-Jy galaxy survey 
on a scale of 7 Mpc, and then evolve this structure linearly back in time
back to $z=50$, following the method proposed by \cite{Kolatt:1996}. We then
use the
resulting field as a Gaussian constraint \citep{Hoffman1991} for
an otherwise random realisation of a flat $\Lambda$CDM model, for which we
assume a present matter density parameter $\Omega_\mathrm{0m}=0.3$, a Hubble constant
$H_0=70$ km~s$^{-1}$~Mpc$^{-1}$ and an root mean square (rms) density fluctuation $\sigma_8=0.9$.  The volume
constrained by the observational data covers a sphere of radius $\sim
110$ Mpc, centred on the Milky Way. This region is sampled with more than 50
million high-resolution dark matter particles and is embedded in a periodic
box $\sim 343$ Mpc on a side. The region outside the constrained volume 
is filled with nearly 7~million low-resolution dark matter particles, allowing 
good coverage of long-range gravitational tidal forces.

Unlike in the original simulation of \citet{Mathis:2002} where only
the dark matter component was present, here we also follow the gas and
stellar components. For this reason we extend the initial conditions by
splitting the original high-resolution dark matter particles into gas and
dark matter particles with masses of $m_\mathrm{gas} \approx 0.69 \times
10^9$~M$_\odot$ and $m_\mathrm{dm} \approx4.4 \times 10^9$~M$_\odot$,
respectively; this corresponds to a cosmological baryon fraction of 13 per
cent. The total number of particles within the simulation is then slightly
more than 108 million and the most massive clusters are resolved by almost
one million particles.  The physics included in the simulation is exactly
the same as that used in the {\it Magneticum Pathfinder} simulation (to
be described in \S\ref{sec_magneticum_pathfinder} below). The lower panel of
Fig.~\ref{fig:foreground} shows the local structures and superclusters
(Perseus-Pisces, Virgo and
Centaurus are all prominent features) and the positions of the known FRBs.
Table~\ref{tab:obs} lists 
DM$_{\rm local~sim}$. for each FRB, defined as the DM contribution along
each sightline from this Local Universe
simulation. 
As can be seen, DM$_{\rm local~sim}$ is relatively small in all cases, with none showing an excess that
corresponds to any specific constrained structures. Since the sightlines are
all through unconstrained regions, the values of DM$_{\rm local~sim}$ have
no specific significance,
and are simply representative of the DM of low-contrast
density enhancements in this local volume. The resulting dispersion from
small-scale structure and the intergalactic medium is incorporated into
the cosmological signal considered in \S\ref{sec_cosmo}, 
and we therefore assume henceforth that the specific contribution to DM
from known structures in the Local Universe is DM$_{\rm LU} = 0$.


\begin{figure*}
\includegraphics[width=0.99\textwidth]{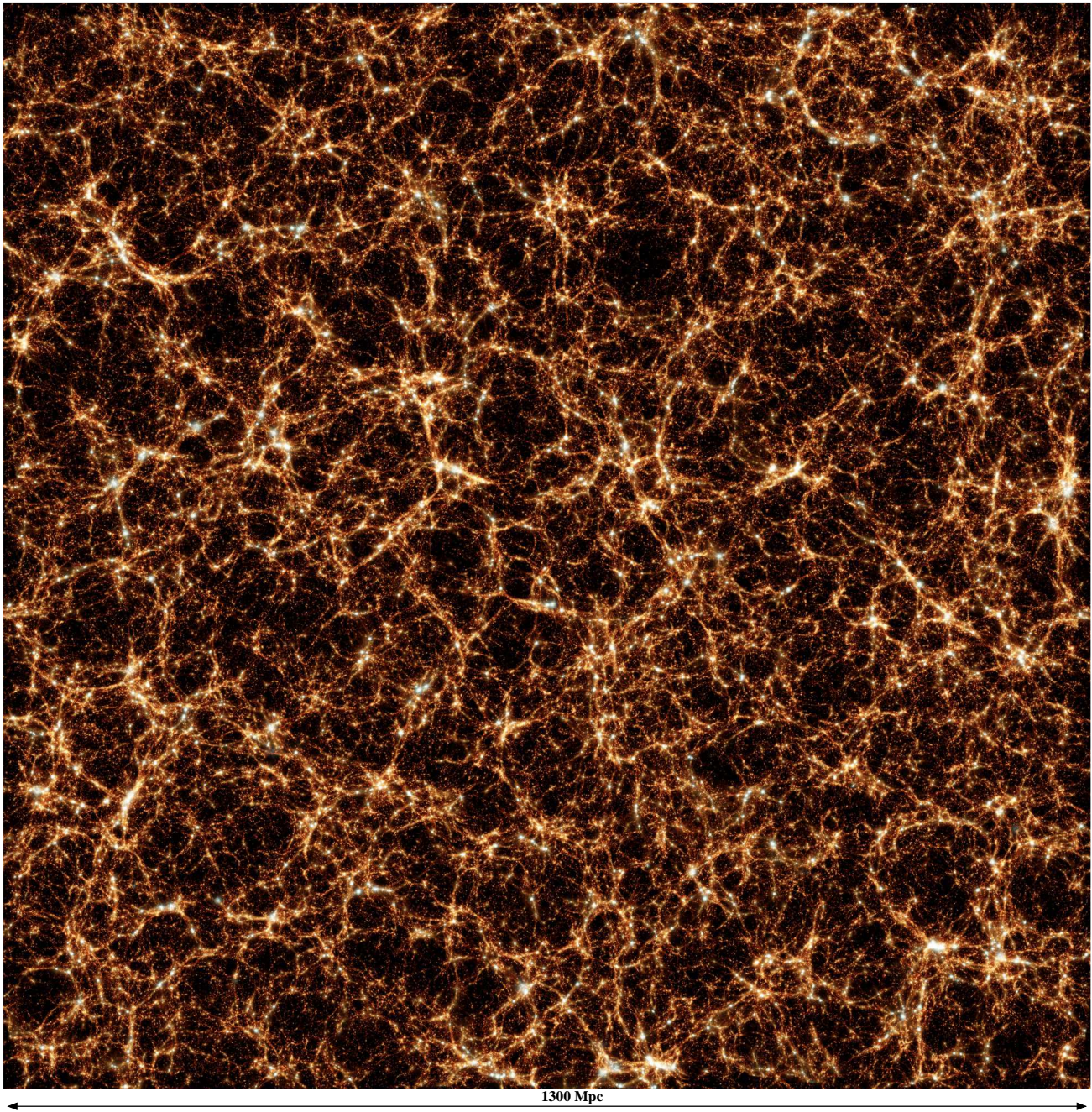}
\caption{The distribution of baryonic material in the largest box of the 
{\it Magneticum Pathfinder} simulation at redshift zero. The colour
indicates the gas temperature, ranging from
dark red (cold) to light blue (hot), combined with the stellar component (white).
The image shows a 100-Mpc thick, 1300-Mpc wide slice through the simulation at $z=0$.}
\label{fig:map}
\end{figure*}

\section{Cosmological Dispersion Measure}
\label{sec_cosmo}

After subtraction of the various foreground contributions to the DM as
defined in Equation~(\ref{eqn_dmsum}) and discussed throughout
\S\ref{sec_signal}, the remaining excess dispersion is listed as DM$_{\rm cosmo}$
in Table~\ref{tab:obs}. In all nine cases, this is by far the dominant
contribution to the total observed DM.

\begin{table*}
\caption{Properties of the two simulation runs from the {\it Magneticum
Pathfinder}\ analysed in this study.}
\label{tab:sims}
\centering
\begin{tabular}{|c c c c c c c c|}
  \hline 
  {Name} &
  {Box size} & {Resolution level} & {Initial particle number} &
  $m_\mathrm{dm}$ & $m_\mathrm{gas}$ & $m_\mathrm{stars}$ & {Softening length (dm, gas, stars)} \\
 & [Mpc/$h$] &  &  & [M$_\odot/h$] & [M$_\odot/h$] & [M$_\odot/h$] &
[kpc/$h$]
 \\ \hline 
1300Mpc/mr & 896 & mr &$ 2 \times 1,526^3$ & $1.3 \times 10^{10}$  & 
$2.6  \times 10^9 $ & $6.5  \times 10^8 $ & 10.0, 10.0, 5.0\\
500Mpc/hr & 352 & hr &$ 2 \times 1,584^3$ & $6.9 \times 10^{8}$  & 
$1.4  \times 10^8 $ & $3.5  \times 10^7 $ & 3.75, 3.75, 2.0\\
\hline
\end{tabular}
\end{table*} 

In the following
sections, we describe our derivation of DM$_{\rm cosmo}$ using hydrodynamic,
cosmological simulations.
As per Equation~(\ref{eqn_cosmo}),
this cosmological contribution to the DM of an FRB is composed of two parts.
The first is DM$_{\rm LSS}$, the signal coming from the diffuse gas within
the cosmic web between the source and the observer. 
The second component is
DM$_{\rm
host}$, the
contribution from the host galaxy of the FRB. Here, due to limited spatial
resolution, our simulations can only capture the hot atmosphere
of virialised halos. For
early-type galaxies, the host contribution should thus be incorporated
by our simulation.
However, for late-type galaxies, the simulation does not capture the
contribution of the gas within the galactic disk. To properly model this
component, additional assumptions or modeling would be required. For the
purposes of the present discussion,
we disregard the disk contribution to DM$_{\rm host}$, because of the likely low
inclination of the disk and because the $(1+z)^{-1}$ dilution of DM$_{\rm host}$
in the observer's frame is larger than for all subsequently encountered dispersive media.
\citep{2013Sci...341...53T,2014ApJ...780L..33M,2014ApJ...788..189G}.

\subsection{The Magneticum Pathfinder}
\label{sec_magneticum_pathfinder}

We use the two largest simulations from the {\it Magneticum
Pathfinder}\footnote{See http://www.magneticum.org.} data set (Dolag et al., in
preparation).
These two simulations use $896h^{-3}$~Mpc$^3$ and $352h^{-3}$~Mpc$^3$ boxes,
simulated using $2\times1526^3$ and $2\times1584^3$ particles, respectively,
where we adopt a WMAP7 \citep{Komatsu11} $\Lambda$CDM cosmology with $\sigma_8 =
0.809$, $h = 0.704$, $\Omega_m = 0.728$, $\Omega_\Lambda = 0.272$,
$\Omega_b = 0.0456$, and an initial slope for the power spectrum of $n_s =
0.963$. A visualisation of a 100-Mpc thick slice from the largest box at
redshift $z=0$
is shown in Fig.~\ref{fig:map}. In 
Table~\ref{tab:sims} we summarise the details of the two simulations,
including the dark matter particle mass, gas particle mass and softening
length. Up to four stellar particles are generated for each gas particle.

Our simulations are based on the parallel cosmological TreePM-smoothed
particle hydrodynamics (SPH) code {\small P-GADGET3}
(\citealp{Springel05gad}). The code uses an entropy-conserving formulation
of SPH \citep{2002MNRAS.333..649S} and follows the gas using a low-viscosity
SPH scheme to properly track turbulence \citep{2005MNRAS.364..753D}.  It
also allows radiative cooling, heating from a uniform
time-dependent ultraviolet (UV) background, and star formation with the associated
feedback processes. The latter is based on a sub-resolution model for the
multi-phase structure of the ISM \citep{Springel03}.

Radiative cooling rates are computed through the procedure
presented by \citet{Wiersma09}. We account
for the presence of the cosmic microwave background (CMB) and
for UV/X-ray background radiation from quasars and
galaxies, as computed by \citet{Haardt01}. The contributions
to cooling from 11 elements (H, He, C, N, O,
Ne, Mg, Si, S, Ca, Fe) have been pre-computed using the publicly
available CLOUDY photoionisation code \citep{Ferland98} for an
optically thin gas in (photo-)ionisation equilibrium. 

In the multi-phase model for star formation \citep{Springel03}, the ISM is
treated as a two-phase medium, in which clouds of cold gas form from
the cooling of hot gas and are embedded in the hot gas phase.
Pressure equilibrium is assumed whenever gas particles are above a given threshold 
density. The hot gas within the multi-phase model is heated by supernovae
and can evaporate the cold clouds. Ten per cent of massive stars 
are assumed to explode as core-collapse supernovae (CCSNe). The energy released
by CCSNe ($10^{51}$~erg per explosion) is modeled to trigger galactic winds
with a mass loading rate proportional to the star-formation rate
(SFR), to obtain a resulting wind velocity $v_{\mathrm{wind}} =
350$~km~s$^{-1}$.
Our simulations also include a detailed model of chemical evolution 
\citep{Tornatore07}. Metals are produced by CCSNe, by Type~Ia supernovae
and by intermediate and low-mass stars in the asymptotic giant
branch (AGB). Metals and energy are released by stars of different mass by
properly accounting for mass-dependent lifetimes (with a lifetime function
as given by \citealp{Padovani93}), the metallicity-dependent stellar
yields of \citet{Woosley95} for CCSNe, the yields of AGB stars from \citet{vandenHoek97} 
and the yields of Type~Ia supernovae from \citet{Thielemann03}. Stars of different
mass are initially distributed according to a \citet{Chabrier03} initial
mass function.

Most importantly, our simulations also include a prescription for black hole
growth and for feedback from active galactic nuclei (AGN).  As for star
formation, accretion onto black holes and the associated feedback is tracked
using
a sub-resolution model. Supermassive black holes are represented by collisionless ``sink
particles'' that can grow in mass either by accreting gas from their environments
or by merging with other black holes. This treatment is based on the model
presented by \citet{Springel05a} and \citet{DiMatteo05} including the same
modifications as in the study of \citet{Fabjan10} plus some further
adaptations \citep[see][for a detailed description]{2014MNRAS.442.2304H}.

We use the {\sc SUBFIND} algorithm \citep{2001MNRAS.328..726S,
2009MNRAS.399..497D} to define halo and sub-halo properties. {\sc SUBFIND}
identifies sub-structures as locally overdense, gravitationally bound groups
of particles. Starting with a halo identified through the Friends-of-Friends
algorithm, a local density is estimated for each particle via adaptive
kernel estimation, using a prescribed number of smoothing neighbours.
Starting from isolated density peaks, additional particles are added in
sequence of decreasing density. Whenever a saddle point in the global
density field is reached that connects two disjoint overdense regions, the
smaller structure is treated as a sub-structure candidate, and the two
regions are then merged.  All sub-structure candidates are subjected to an
iterative unbinding procedure with a tree-based calculation of the
potential. These structures can then be associated with galaxies, and their
integrated properties (such as stellar mass or star-formation rate) can then
be calculated.
Note that with an adopted resolution
limit for our simulations of $3\times10^9$~M$_\odot$, any detected galaxy
is assumed to contain a central supermassive black hole (SMBH).

\begin{figure*}
\includegraphics[width=0.49\textwidth]{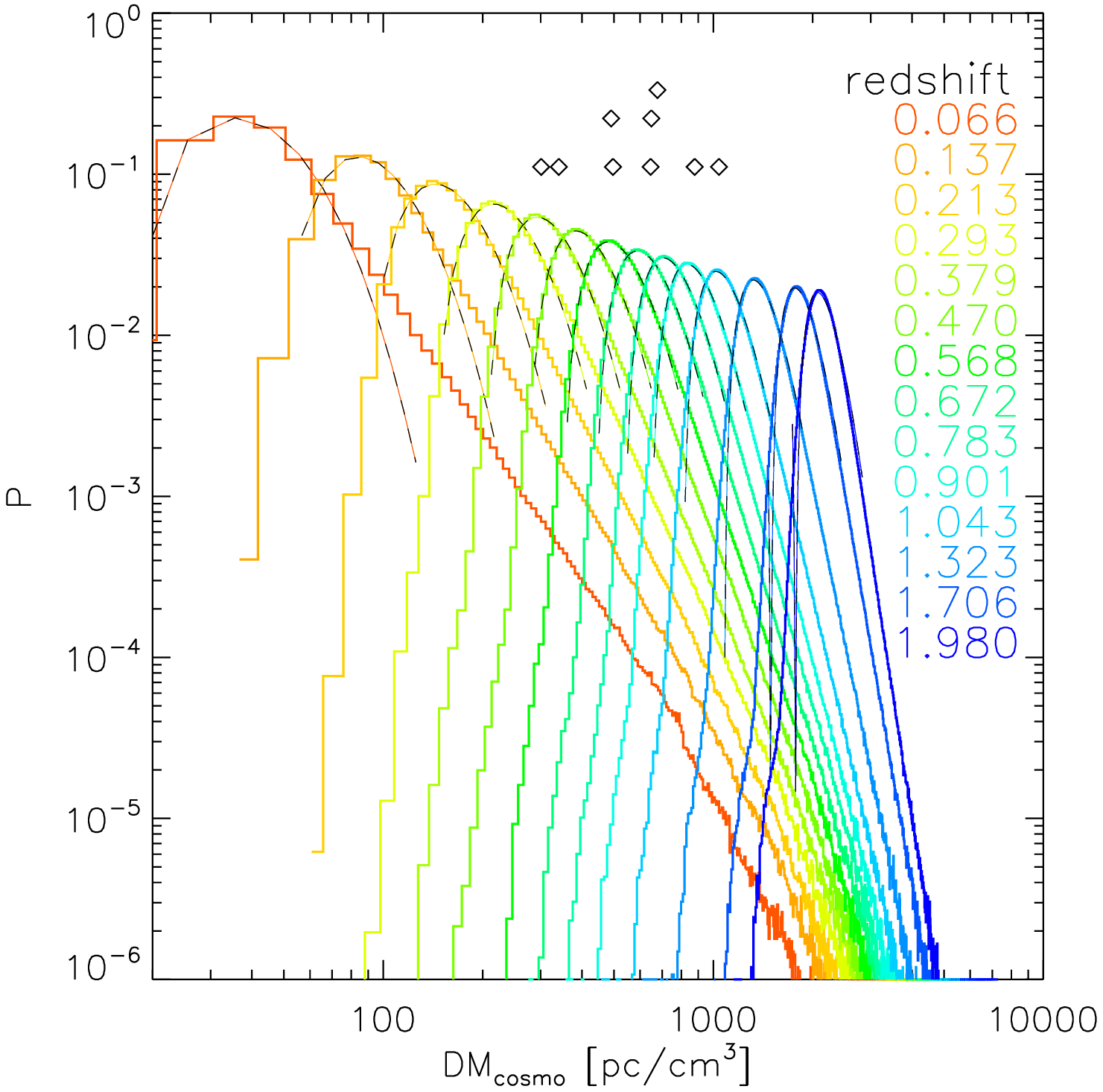}
\includegraphics[width=0.49\textwidth]{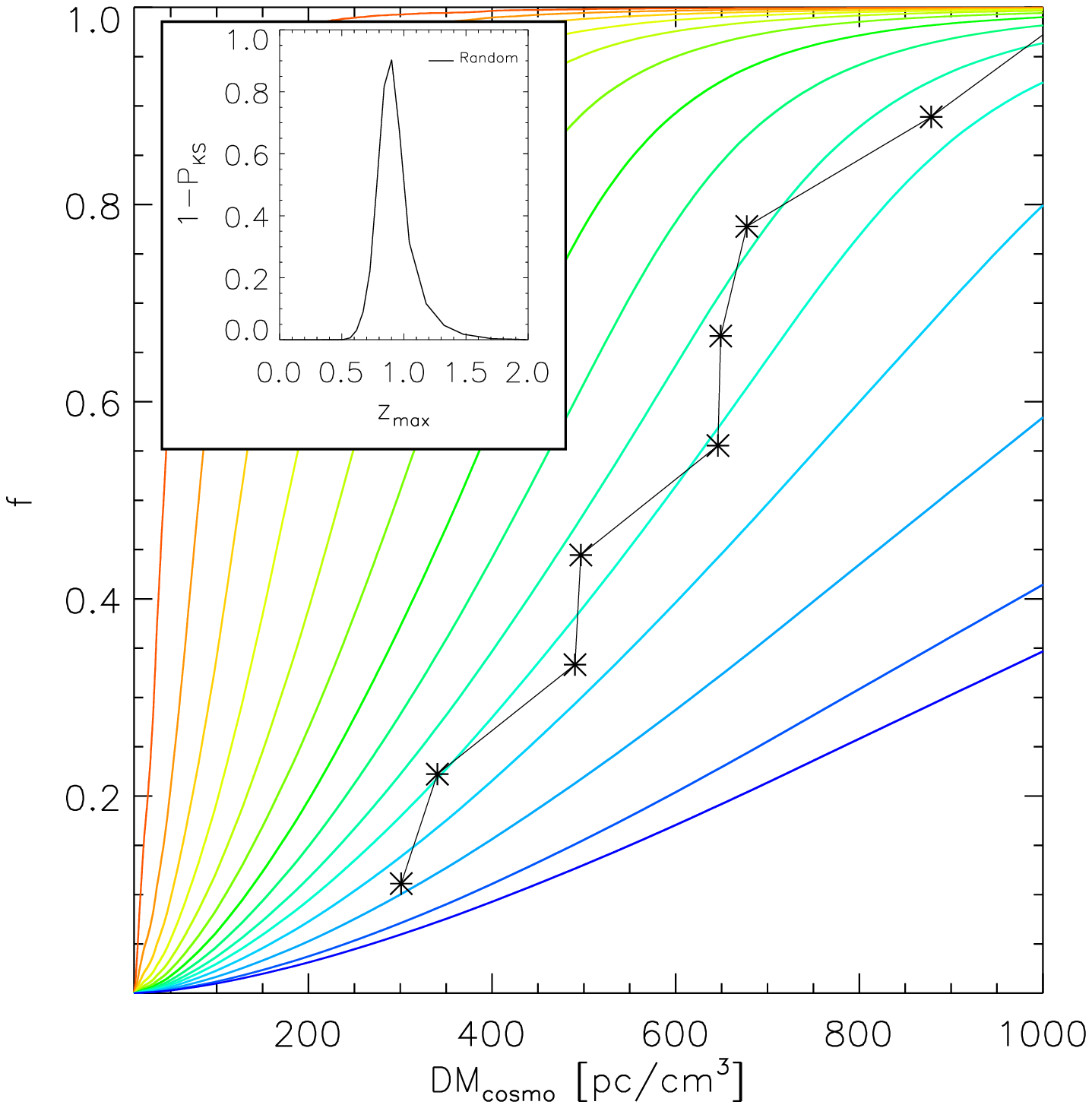}
\caption{The expected distribution of  DM$_{\rm cosmo}$ for FRBs derived from the
medium-resolution simulation.
The left panel shows histograms of DM$_{\rm cosmo}$ as a function of
redshift from the {\it 1300Mpc/mr} simulation,
with each coloured curve representing the distribution
of DM$_{\rm cosmo}$ for a large ensemble
of sightlines at a fixed redshift. Values of DM$_{\rm
cosmo}$ observed for FRBs in Table~\ref{tab:obs} are shown by black
diamonds.
Fits to the simulated distributions of DM$_{\rm cosmo}$
are shown as black dashed lines (see text for details).
The right panel shows cumulative distributions of DM$_{\rm cosmo}$,
obtained by integrating
light-cones through the hydrodynamic simulation up to a maximum redshift
$z_{\rm max}$,
and assuming that FRBs are randomly distributed in the Universe --- i.e.,
distributed randomly within the co-moving volume of every slice of the
light-cone. Each coloured curve shows the cumulative distribution of
DM$_{\rm cosmo}$ for a different redshift $z_{\rm max}$, using the same colours
as in the left-hand panel.
The black points show the corresponding 
cumulative distribution of DM$_{\rm cosmo}$ for FRBs from  Table~\ref{tab:obs} (noting that DM$_{\rm
cosmo}$ for FRB~110703 sits just outside the DM range plotted).  The
inset shows the probability derived from a KS test that the simulated and
observed distributions of  DM$_{\rm cosmo}$ could be drawn from the same
underlying distribution, as a function of $z_{\rm max}$.
}
\label{fig:dms_universe}
\end{figure*}

\subsection{Calculating the Cosmological Dispersion Measure}
\label{sec_calc}

Within the simulation we assume a primordial mixture of hydrogen
and helium with a hydrogen mass fraction of 0.752, and when calculating the
electron density we take into account the actual ionisation state
of the medium. We disregard star-forming particles in this calculation,
since their multi-phase nature means that their free electron density is
not properly characterised. The cosmological frequency shift is taken into
account when integrating the free electron density 
\citep{2014ApJ...780L..33M,2014ApJ...783L..35D}, such that:
\begin{equation}
\mathrm{DM}_{\rm cosmo} (z_\mathrm{max}) = \int_0^{z_\mathrm{max}}
\frac{n_e(z)}{1+z}~dl,
\label{eqn_cosmo_dm}
\end{equation}
where we integrate up to some maximum redshift of interest
$z_\mathrm{max}$.

To actually construct past light-cones from the simulations, we follow
the common approach to stack the co-moving volumes (e.g. placing them
at the proper distance $w(z_i)$) of the simulations
(see for example \citet{2007MNRAS.378.1259R,2010ApJ...721...46U}
for similar approaches). To avoid replications of similar structurs
we randomized cosecutive slices by rotating and shifting. Our simulation
volumes however are big enough so that we do not need to duplicate
the simulation volumes for our 36 different individual slices.
The number of outputs produced in the simulation was chosen 
so that the required radial integration length ($w(z_{i+0.5})-w(z_{i-0.5})$) of the
individual slices always fit entirely within the simulated volume, which
is placed at the proper distance $w(z_i)$ according to the assumed cosmology.
The opening angle is chosen so that the orthogonal extent --- which
depends on the angular diameter distance at the redshift of a slice -- -
always fits entirely within the simulated volume. The cosmological
signal up to a maximum redshift $z_\mathrm{max}$ is thus approximated by
the stacking of the individual slices:

\begin{equation}
\mathrm{DM}_{\rm cosmo} (z_\mathrm{max}) = \sum_{i=0}^{i_{\rm max}} \int_{w(z_{i-0.5})}^{w(z_{i+0.5})}
\frac{n_e(l)}{1+z_i}~dl.
\label{eqn_cosmo_dm_sum}
\end{equation}

We produce maps of the integrated electron density across each
slice using {\sc SMAC} \citep{dolag2005}, each resolved with $4096\times4096$
pixels and covering a field of view of $13^\circ\times13^\circ$ for the
{\it 1300Mpc/mr} simulation and $5^\circ\times5^\circ$ for the {\it 500Mpc/hr}
simulation. Each of these slices represents the corresponding cosmological
volume within the light-cone, while the actual integration is done converting the simulation to physical units first. We are using this volume, as well as the
galaxies (and their properties) to weight the our source models as introduced
in section \S\ref{sec_high}.

\begin{figure*}
\includegraphics[width=0.99\textwidth]{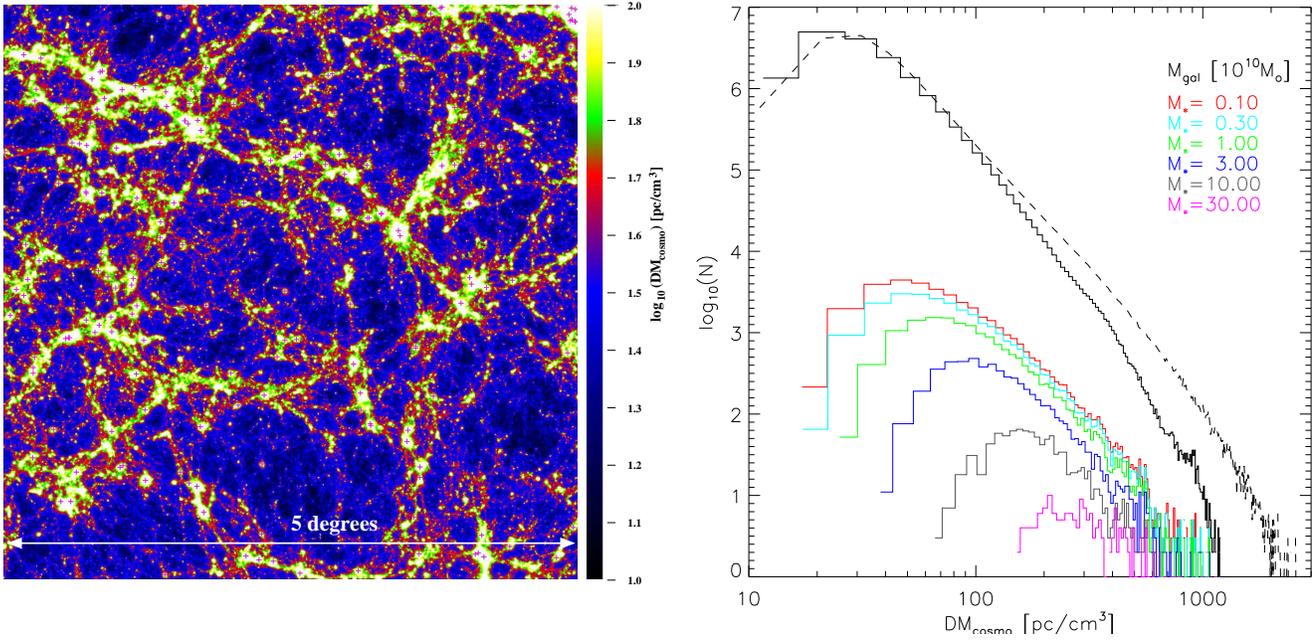}
\caption{Results from the high-resolution simulation. The left-hand panel
shows the distribution of DM$_{\rm cosmo}$ in  the {\it 500Mpc/hr}
simulation in a roughly 160-Mpc-thick slice of the light-cone 
at $z=0.5$; pink crosses mark the positions of the most massive galaxies.
In the right-hand panel, we show the
corresponding overall distribution function for DM$_{\rm cosmo}$ (black
line), and the restricted distribution of DM$_{\rm cosmo}$ only at the positions of galaxies
(coloured lines, with different colours showing differnet minimum thresholds
for stellar massss).  For comparison, the distribution of
DM$_{\rm cosmo}$ for the lower-resolution but larger {\it 1300Mpc/mr}
simulation is shown by a dashed line. With the exception of the tail at
extreme DMs, the distributions of DM$_{\rm cosmo}$ in the two simulations
are reasonably similar.
}
\label{fig:dm_slice}
\end{figure*}

\subsection{Medium-Resolution Simulation}
\label{sec_med}

We first consider results from the {\it 1300Mpc/mr} simulation, which covers
a large cosmological volume at intermediate spatial resolution. This allows
us to study the overall distribution of DM$_{\rm cosmo}$ down to the scale
of galaxy groups.
The resulting distribution of DM$_{\rm cosmo}$ as a function of
$z_{\rm max}$
can be seen in the left panel of Fig.~\ref{fig:dms_universe}, where we
show the distribution of DM$_{\rm cosmo}$ in the light-cone when integrating
up to the indicated redshift.
The nine observed values of DM$_{\rm cosmo}$ listed in Table~\ref{tab:obs} are shown
as black diamonds in Fig.~\ref{fig:dms_universe}, 
confirming the conclusions of previous authors that
FRBs occur at 
large cosmological distances, $0.5 \la z \la 1$
\citep{2003ApJ...598L..79I,2004MNRAS.348..999I,2007Sci...318..777L,2013Sci...341...53T}.

The dashed black lines shown in the left-hand panel of
Fig.~\ref{fig:dms_universe} indicate fits to the DM$_{\rm cosmo}$ distribution $P({\rm DM}_{\rm cosmo})$
of the form:
\begin{eqnarray}
   P({\rm DM}_{\rm cosmo})=A(z) &\times& \left[{\rm DM}_{\rm cosmo}+{\rm DM}_{00}(z)\right]^2 \nonumber \\
                             &\times& \mathrm{exp}\left(-\frac{{\rm DM}_{\rm como}+{\rm DM}_{01}(z)}{\sigma(z)}\right)
\end{eqnarray}
where the constants can be written as function of redshift,
\begin{equation}
A(z)=2\times10^{-11}z^{-9}+1.3\times10^{-7}z^{-4}+5\times10^{-7}(z-2.325)^2,
\end{equation}
\begin{equation}
{\rm DM}_{00}(z)=-770z^{1.3}+31.2\sqrt{z},
\end{equation}
\begin{equation}
{\rm DM}_{01}(z)=-108+926z+67z^3,
\end{equation}
\begin{equation}
\sigma(z)=126.9z+2.03-27.4z^2.
\end{equation}
Note that we only fit the distribution of DM$_{\rm cosmo}$ down to 1\% of the
maximum value for each value of $z$, since this captures the bulk of the
signal and because at their highest values the distributions of DM$_{\rm
cosmo}$ take on a complex shape that can only be fit properly with a
combination of several broken power laws. Note also that this tail toward
large values of DM$_{\rm cosmo}$ is sensitive both to the largest
supercluster structures present (and therefore to the size of the underlying
cosmological simulation) and to the ability of the simulation to properly
capture the inner regions and cool cores of galaxy clusters (see also the
left panel of Fig.~\ref{fig:dm_slice} and the associated discussion below). 

The right-hand panel of Fig.~\ref{fig:dms_universe} shows the expected
cumulative distribution of DM$_{\rm cosmo}$ if the detected FRBs originate
from random
locations throughout the Universe  (i.e., at a probability of incidence
proportional to the co-moving volume within the light-cone) 
up to some maximum observable redshift $z_{\rm max}$.
The observed cumulative distribution of DM$_{\rm cosmo}$ for FRBs, shown in
black, closely matches the simulated distribution for $z_{\rm max} \sim
1$.
We quantify this in
the inset to the right-hand panel of Fig.~\ref{fig:dms_universe}, where we
show the
result of a series of
Kolmogorov-Smirnoff (KS) tests between the observations and the simulated
distributions as a function of $z_{\rm max}$. This demonstrates that the data
only match the simulations in a narrow range around $z_{\rm max} \approx 0.9$.

We emphasise that although the {\it 1300Mpc/mr} simulation is currently the
largest of its kind, the corresponding spatial resolution is not good enough
to properly capture details in the halo structure (and especially 
precise stellar components of halos) at scales below massive galaxies.
We therefore only use the {\it 1300Mpc/mr} simulation for the overall distribution
of DM$_{\rm cosmo}$ (i.e., the component dominated by DM$_{\rm LSS}$), a
regime for which the {\it 1300Mpc/mr} simulation is superior to the
higher-resolution {\it 500Mpc/hr} simulation
because the former captures structures on 
much larger scales. 
We switch to the smaller, higher-resolution
{\it 500Mpc/hr} simulation when stellar properties of the halos get
important (e.g., when trying to account for DM$_{\rm host}$),
as we next consider in~\S\ref{sec_high}.

\begin{figure*}
\includegraphics[width=0.49\textwidth]{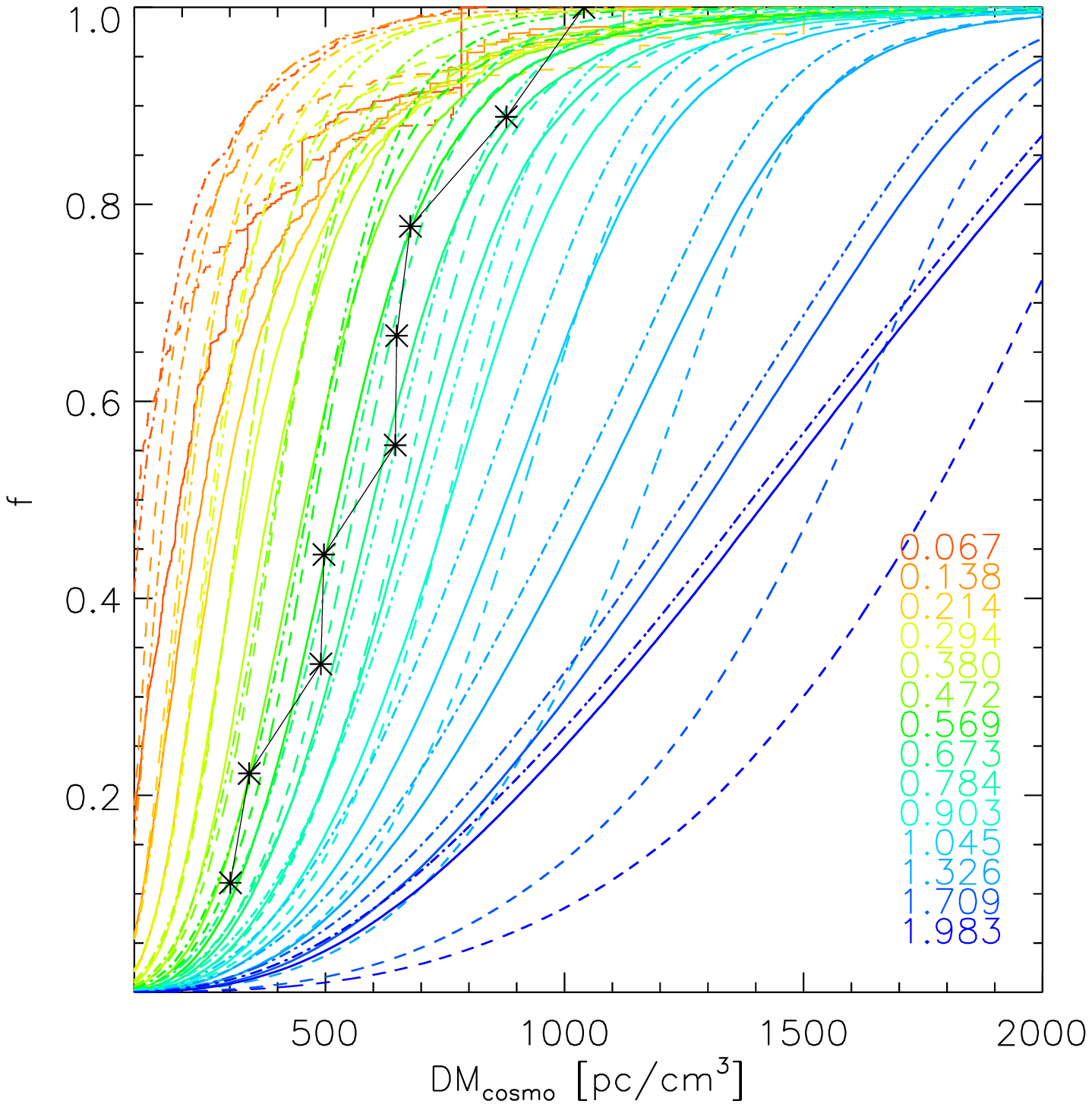}
\includegraphics[width=0.49\textwidth]{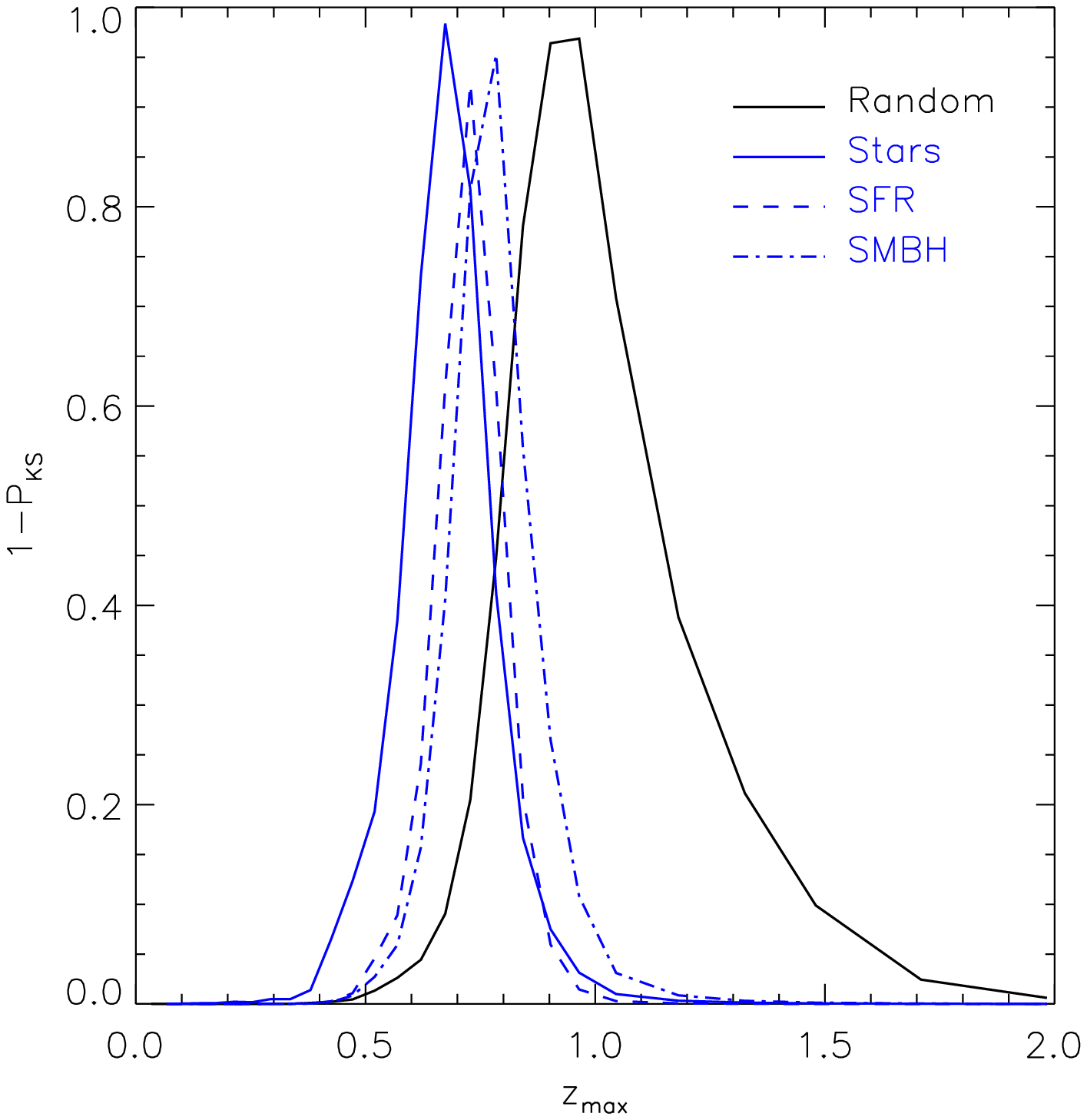}
\caption{The expected cumulated distribution of  DM$_{\rm cosmo}$ as a
function of maximum observed FRB redshift, assuming that FRBs only occur in
galaxies, and that the FRB rate for a given galaxy depends on stellar mass,
(solid line), star-formation rate (dashed line) or the presence of a central
SMBH (dash-dotted line). The left-hand panel shows the
simulated cumulative distributions of DM$_{\rm cosmo}$ as a function of
maximum redshift, $z_{\rm max}$, for these three models, with the observed distribution
overlaid in black.  The right-hand panel shows the results of KS tests
between each model and the data, each as a function of $z_{\rm max}$.
} 
\label{fig:dms_source} 
\end{figure*}

\subsection{High-Resolution Simulation}
\label{sec_high}

In \S\ref{sec_med}, we used the medium-resolution 
{\it 1300Mpc/mr} simulation to determine the distribution of DM$_{\rm
cosmo}$ for FRBs distributed randomly over the cosmic volume out to some
maximum observable redshift. 
We now use the  high-resolution {\it 500Mpc/hr} simulation to calculate the
corresponding distribution of DM$_{\rm cosmo}$ when FRBs are embedded in
potential host galaxies.
We consider three simple models in which the spatial distribution of detectable
FRBs within the simulation volume is correlated with
the properties and locations of individual galaxies:
\begin{enumerate}
\item FRBs trace the total stellar component, as might result if FRBs are
produced by an evolved population such as merging neutron stars. In this
case, we assume that the FRB rate is proportional to the stellar mass within
each galaxy. To compute the spatial distribution of FRBs, we only allow FRBs
to occur at pixels in our simulation at which we identify a galaxy, and we
weight the rate of FRB occurrence by the corresponding stellar mass.

\item FRBs are associated with massive stars, as would result if FRBs
are produced by supernovae or young neutron stars. We here assume
that the FRB rate is proportional to the current star-formation rate.
We calculate the rate of FRBs by again only  allowing an FRB to occur
at a pixel associated with an individual galaxy, but we now weight
the FRB rate
by the current star-formation rate of that galaxy.

\item FRBs are associated with activity or interactions around the
SMBH in a galaxy's nucleus.  
Again we assume FRBs can only occur at pixels associated with individual
galaxies, but we give all such pixels equal weight. 
\end{enumerate}

In each case, we wish to compute the expected distribution of DM$_{\rm
cosmo}$, and compare this to the FRB observations given in
Table~\ref{tab:obs} to see if we can discriminate between possible FRB
mechanisms.  We construct simulated distributions of DM$_{\rm cosmo}$ by
identifying the position of galaxies within each slice of the light-cone.
The predicted value of DM$_{\rm cosmo}$ for an FRB occurring within a given
galaxy (disregarding the disk contribution as discussed in
\S\ref{sec_cosmo}) is then the contribution within the light-cone up to the
position of that galaxy, as per Equation~(\ref{eqn_cosmo_dm}).  Using the
global properties of this galaxy determined as described in
\S\ref{sec_magneticum_pathfinder}, we then assign a relative probability for
the occurrence of an FRB at this location by weighting the correspondingly
calculated value of  DM$_{\rm cosmo}$ as per one of the three schemes
described above.  

The left-hand panel of Fig.~\ref{fig:dm_slice} shows the contribution to
DM$_{\rm cosmo}$ in the light-cone from a slice at redshift $z\approx0.5$ in
the {\it 500Mpx/hr} simulation, overplotted with the positions of the most
massive galaxies (which correspond to the central galaxies in groups and
clusters). The right-hand panel of Fig.~\ref{fig:dm_slice}  shows the
distribution functions for DM$_{\rm cosmo}$ through all pixels in the
full slice (black line) and only for pixels
associated with galaxies above a
threshold in stellar mass (colour coded).  We also show the
distribution of DM$_{\rm cosmo}$ for the same slice in the {\it 1300Mpc/mr}
simulation (dashed line), in order to
illustrate the contribution of even larger structures that are only present in the
larger cosmological box. These large structures manifest as an excess
in the tail of the distribution 
at the largest values of DM$_{\rm cosmo}$. However, for the range
in DM$_{\rm cosmo}$ corresponding to halos of all masses,
the overall distributions of DM$_{\rm cosmo}$ for the medium- and
high-resolution simulations are reasonably similar and the differences due
to resolution and box size will not significantly alter any of our conclusions.
In general, the range of values derived for DM$_{\rm cosmo}$
agrees with the estimates made by \citet{2014ApJ...780L..33M}.  Note that the different
models for the spatial locations of FRBs all utilise the same underlying
distributions in DM$_{\rm cosmo}$, but give different weights to the
individual pixels depending on the global properties of the galaxy associated with each
pixel.

We perform this calculation for every slice of our light cone and thereby
construct distributions of DM$_{\rm cosmo}$ as a function of
$z_{\rm max}$ for each of the three different weighting schemes described
above. The resulting cumulative distributions of DM$_{\rm cosmo}$ are shown
in the left-hand panel of Fig.~\ref{fig:dms_source}, together with the
corresponding observed cumulative distribution of DM$_{\rm cosmo}$ for the FRBs listed
in Table~\ref{tab:obs}.

\begin{figure*}
\includegraphics[width=0.49\textwidth]{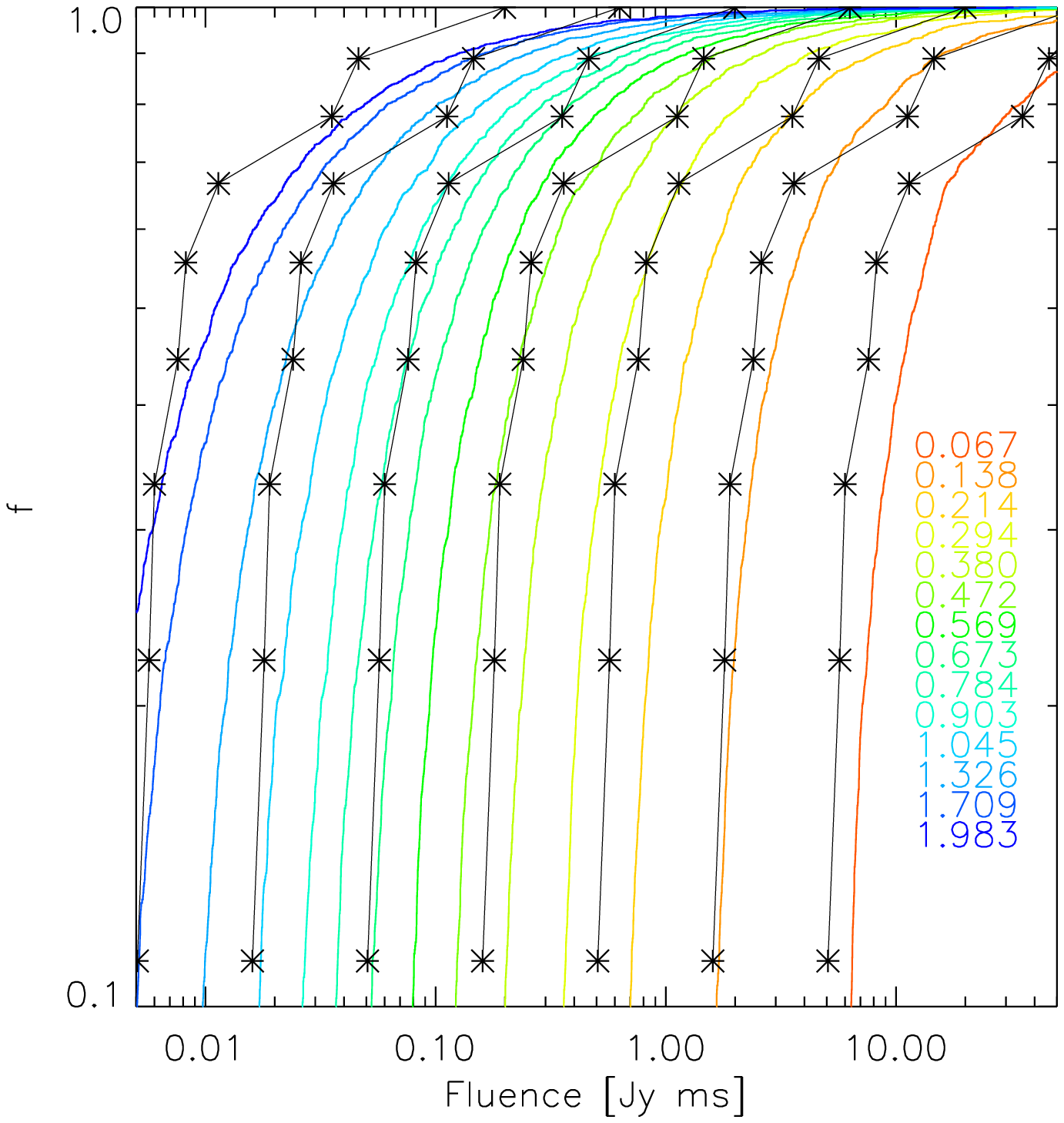}
\includegraphics[width=0.49\textwidth]{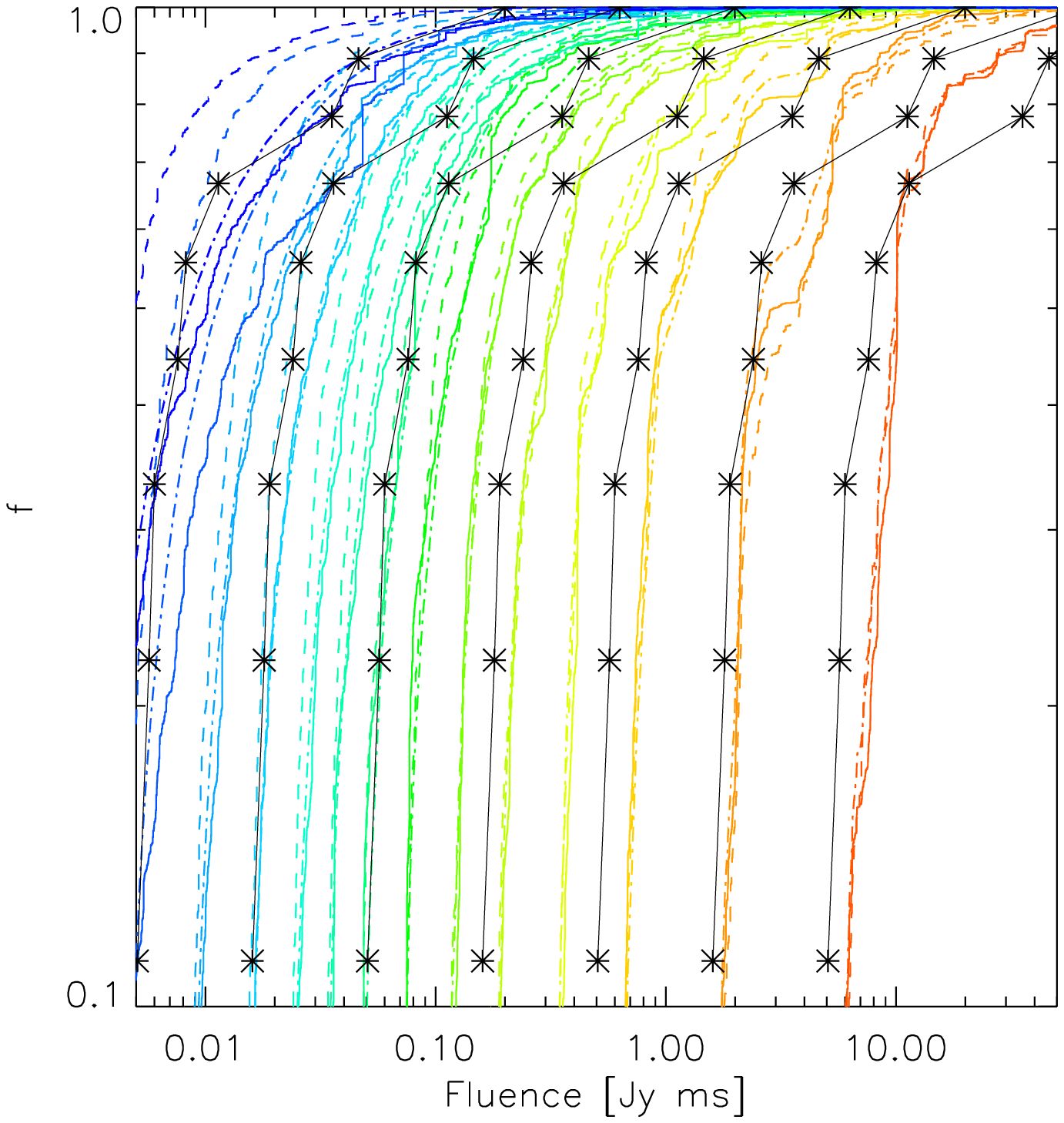}
\caption{The expected distribution of FRB fluences as a function of maximum
redshift, $z_{\rm max}$, assuming a nominal energy $E = 1.4\times10^{39}$~erg
in the radio pulse.
The left-hand panel shows simulations of the cumulative
distribution of fluence as a function of maximum redshift  for randomly
distributed FRBs, 
while the right-hand panel shows the equivalent calculations when
FRBs are associated with host galaxies and when
the FRB rate depends on total stellar mass (solid line),
star-formation rate (dashed line) or the presence of a central black hole (dotted
line). The colours correspond to a range of maximum redshifts as per the
legend to Fig.~\ref{fig:dms_universe}.  Observed FRB fluences are
overplotted in black: the second curve from the right shows the fluences in
Table~\ref{tab:obs}, while the other six curves show fluences
shifted up or down in successive steps of half a decade to provide
a scaling to other values of $E$.}
\label{fig:lum_vol}
\end{figure*}

There are only nine observed data points at present, collected in a very
inhomogeneous way rather than from a single survey. Thus we cannot draw
robust conclusions from the present sample.  Nevertheless, the left-hand
panel of Fig.~\ref{fig:dms_source} shows clearly that the detected FRBs
extend up to a maximum redshift $z_{\rm max} \approx 0.6-0.9$, independent of their
origin or detailed spatial distribution with respect to any host galaxies.
We quantify this in the right-hand panel of  Fig.~\ref{fig:dms_source},
where we show the results of KS tests between the observed values of
DM$_{\rm cosmo}$ and the simulated distributions as a function of maximum
redshift for the three scenarios involving FRBs in host galaxies, and also
for the random (unweighted) distribution of FRBs considered in
\S\ref{sec_med}.\footnote{Note that 
the difference in the cosmological
DM signal between the medium- and high-resolution simulations is very small
and does not affect our conclusions, as can be seen
comparing the inset in Fig.~\ref{fig:dms_universe} with the corresponding
black line in the right panel of Fig.~\ref{fig:dms_source}.}
It is clear that all four possibilities are a reasonable
match to the observations, although the best-fitting values of $z_{\rm max}$
are smaller for FRBs in host galaxies than for FRBs distributed randomly (as
expected given that the former involves sightlines biased toward high DMs as
per Fig.~\ref{fig:dm_slice}). 
The differences in the best fits for $z_{\rm max}$ for the three host-galaxy
models are small, but the fact that case (i) gives a lower redshift than
case (ii), and that case (ii) is lower than case (iii), can be understood
qualitatively. If we adopt case (i) in which the frequency of FRBs tracks
stellar mass, many
sightlines to FRBs then
involve high-mass systems, which therefore contribute more to DM$_{\rm
host}$ and hence reduce the required value of $z_{\rm max}$.
However, massive galaxies experience reduced star-formation due to SMBH activity
in these systems; thus in case (ii), FRBs occur on sightlines to lower-mass
galaxies, such that DM$_{\rm host}$ is lower and hence $z_{\rm max}$ is
higher. Finally, FRBs associated with SMBHs result in a set of sightlines
that have no weighting at all for the size of the host galaxy, and
therefore result in even larger values of $z_{\rm max}$.

The similarities between the form of the cumulative DM$_{\rm cosmo}$
distributions of different weighting schemes at these redshifts (seen as the
green curves in the left panel of Fig.~\ref{fig:dms_source}) means that even
once we obtain a much larger sample of FRBs, it will be difficult to use
their DMs to discriminate between different possible origins if the observed
values of DM$_{\rm cosmo}$ continue to mostly fall in the range
$\sim300-1000$~pc~cm$^{-3}$. Conversely, as we approach $z_{\rm max} \sim 2$
the distributions of  DM$_{\rm cosmo}$ diverge for the different weighting
schemes (seen in blue in the left panel of Fig.~\ref{fig:dms_source}).
The much higher star-formation rate at these earlier epochs leads to
larger differences in the weighting schemes between models, and thus to 
a broader range in the predicted distributions of DM$_{\rm cosmo}$.
If (through either improved sensitivity or simply a larger sample size) we can begin to detect an appreciable fraction of FRBs
with DM$_{\rm cosmo} \ga 1000-1500$~pc~cm$^{-3}$,
we may be able to
distinguish between the different models, especially between FRB mechanisms
that track the star-formation rate compared to other possibilities.

\section{Expected Fluences and Energies}
\label{sec_fluence}

We now consider the implications for the observed distribution of FRB
fluences.
We consider the same three possible spatial
distributions for FRBs as discussed in \S\ref{sec_high}, along with a uniform (random) distribution of FRBs
within the co-moving volume as in \S\ref{sec_med}.
However, while before we calculated distributions of DM from the
simulations, here we infer fluences, and use these to test the hypothesis
that FRBs are standard candles, each with the same emitted radio energy.

We assume that each FRB has the same isotropic 
total energy, $E$, where:
\begin{equation}
\frac{E}{4\pi D_\mathrm{lum}^2(z_0)} = F_{\nu_0}~\nu_0 (1+z_0)~,
\label{eqn:lumdist}
\end{equation}
for an FRB of received fluence $F_{\nu_0}$ at an observing frequency
$\nu_0$, where $D_\mathrm{lum}$ is the luminosity distance for the FRB's
redshift, $z_0$.  

\begin{figure}
\includegraphics[width=0.49\textwidth]{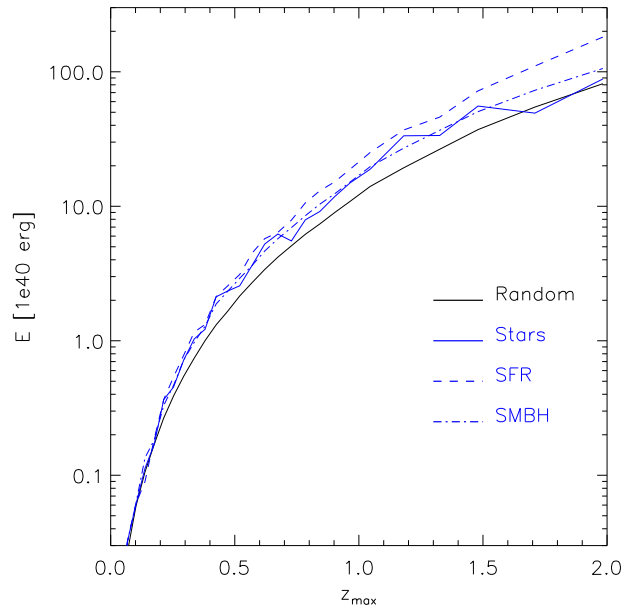}
\caption{The isotropic FRB energy needed to match the observed distribution
of cumulative fluences as a
function of maximum redshift and for different models for the spatial
distribution of FRBs, assuming all FRBs to be standard candles and using
Equation~(\ref{eqn:lumdist}) to convert between observed fluence and
isotropic energy.
}
\label{fig:lum_power}
\end{figure}

Fig.~\ref{fig:lum_vol} shows the expected cumulative
distributions in fluences for a range of values of $E$ and as a function of
maximum FRB redshift: the
left panel shows how such fluences should be distributed if FRBs are located
randomly in the cosmic volume (see \S\ref{sec_med}), while the right panel shows the corresponding
distribution if the number of FRBs per galaxy scales as  the total stellar
mass, as the star-formation rate, or at an equal rate per massive galaxy
(see \S\ref{sec_high}).
For comparison, the cumulative distribution of observed fluences from
Table~\ref{tab:obs} is overplotted as black points, and also shifted up or
down to mimic different values of the fiducial energy that we assign to the
FRBs.
Again having only nine data points does not allow robust conclusions, but
overall the shapes of the predicted distributions in fluence are all similar to
that observed, showing that the data at present are consistent with the FRB
population being standard candles. This favours mechanisms that produce
FRBs through a deterministic process with a small number of free
parameters \citep[e.g., blitzars or neutron-star
mergers;][]{Totani2013,Falcke2014} over stochastic
process such as flares or reconnection events that typically show a wide
distribution of energies
\citep{Popov2013,2014MNRAS.442L...9L,2014MNRAS.439L..46L}.
The fluences for a uniform (random) distribution show a very good
agreement with the data. Models in which FRBs
trace galaxies provide a slightly poorer match to the observed fluences, but 
this difference is not statistically significant, and none of
the simulated distributions can be excluded.
In Fig.~\ref{fig:lum_power} we show the value of the isotropic energy
required to match the distribution of observed fluences as a function of maximum observed
redshift, under the assumption that FRBs are
standard candles, and for the same four models for the spatial
distribution of FRBs as shown in Fig.~\ref{fig:lum_vol}. Note that this only depends on the
spatial distribution of possible FRB hosts in the simulations. 
With the DM distributions considered in \S\S\ref{sec_med} \&
\ref{sec_high},
we can then infer $E \sim 7\times10^{40}$~erg for any of the three
models in which FRBs trace large-scale structure, for which we found
$z_{\rm max} \sim 0.6-0.7$ in Fig.~\ref{fig:dms_source}. 
For randomly distributed FRBs, for which 
we found $z_{\rm max} \approx 0.9$ in \S\ref{sec_med}, $E$ is slightly larger than this (
$E \sim 9\times10^{40}$~erg).
We note that these energies, inferred using Equation~(\ref{eqn:lumdist}),
are 1--2 orders of magnitude larger than those reported in most other papers
\cite[e.g.,][]{2013Sci...341...53T,2015MNRAS.447.2852K}, because these previous
authors have omitted the factor $4\pi$ needed to calculate isotropic
energies, and have multiplied by the observing bandwidth rather than the
observing frequency. The latter is far less dependent on the specifics of
the observations, and (in the absence of any spectral index information) is
a better rough estimate of the integrated radio energy of the burst;
see \S3.1 of \citet{2014ApJ...797...70K} for a more detailed treatment.
The energies we have calculated are a very good match to the predicted energy
release $E \approx 3 \times 10^{40}$~erg predicted by the blitzar model for
FRBs presented by
\citet{Falcke2014} and also agree with $E \ga 10^{40}$~erg expected from
FRBs generated by magnetar flares \citep{2014MNRAS.442L...9L,2014ApJ...797...70K}.

We note from Table~\ref{tab:obs} that FRBs~110220 and 140514 occurred at
almost the same position on the sky,\footnote{This is not a complete coincidence:
\citet{2014arXiv1412.0342P} discovered FRB~140514 while searching for repeat
emission from FRB~110220.}
but with different DMs and different
fluences. If FRBs are standard candles, the simplest expectation along a
single sightline is that an FRB with a higher DM is farther away and thus
must have a lower fluence. This is the opposite to what is observed for
these two sources, for which FRB~140514 has a lower observed DM than
FRB~110220 but also a lower fluence.
However, given the significant angular structure  seen
in the simulation (see Fig.~\ref{fig:dm_slice}), we cannot exclude large
fluctuations in DM, even along sightlines separated by 0.1~degrees.
In addition, the reported fluences of FRBs have significant systematic
uncertainties due to their unknown location within the telescope beam
\citep[see][]{2014arXiv1404.2934S,2015MNRAS.447.2852K}.

\section{Conclusions}

We have used a set of advanced cosmological hydrodynamic simulations to
investigate the contributions to dispersion measures of fast radio bursts
from the Milky Way disk and halo, from the local Universe, from cosmological
large-scale structure, and from potential host galaxies. Through this
combination of calculations, we have made predictions for the expected DMs
of FRBs distributed over different redshift ranges and for differing spatial
distributions.

For the foreground (non-cosmological) contributions to DM, we obtain two
main results:
\begin{enumerate}
\item The Milky Way's hot halo contributes an additional
$\sim$30~pc~cm$^{-3}$ to the total DM, over and beyond the DM contribution
from the Galactic disk predicted by the NE2001 model of
\citet{2002astro.ph..7156C,2003astro.ph..1598C}. Except for FRBs at low
Galactic latitudes ($|b| \la 20^\circ$), this means that the full Galactic
contribution to FRM DMs is approximately double previous estimates.  

\item By using a constrained simulation of the local Universe, we exclude any
significant contribution to FRB DMs from prominent structures out to a
distance of $\sim110$~Mpc.
\end{enumerate}

From our simulations of DMs at cosmological distances, we can make four
additional conclusions:
\begin{enumerate}
\setcounter{enumi}{2} 
\item The observed DM distribution for the available sample of nine extragalactic
FRBs is consistent with a cosmological population detectable
out to a redshift $z_{\rm max} \approx 0.6-0.9$,
regardless of the specifics of how FRBs are distributed with respect to 
large-scale structure or the properties of their host galaxies.

\item If future observations can extend the FRB population to higher DMs and
higher redshifts (DM$_{\rm cosmo} \ga 1000-1500$~pc~cm$^{-3}$, $z_{\rm max}
\approx 2$) than for the currently known sample, we will be able to use the
resulting DM distribution to determine whether FRBs are related to recent
star formation or have some other origin.

\item The distribution of observed FRB fluences is consistent with a
standard-candle model, in which the radio emission from each FRB corresponds
to the same isotropic energy release.

\item Under the assumption that FRBs are standard candles, the isotropic
energy associated with each radio burst is $\sim7\times10^{40}$~erg if FRBs
are embedded in host galaxies and trace large-scale structure, or
$\sim9\times10^{40}$~erg if FRBs occur at random locations in the Universe.
\end{enumerate}
The blitzar model, in which a supramassive young neutron star is initially
supported against gravitational collapse through its rapid rotation, but
then later implodes to form a black hole once it has spin down sufficiently
\citep{Falcke2014}, is an FRB  mechanism that may meet this joint
requirement that FRBs are standard candles and that the radiated energy is
$\ga10^{40}$~erg. A more statistically robust distribution of fluences
resulting from additional FRB detections will be able to better test whether
FRBs are indeed standard candles, while an extension to higher redshifts can
test whether FRBs track the star-formation rate as expected for the blitzar
model.

There are many additional issues that we have not considered in this initial study.
From an observational perspective, the fluxes and fluences of FRBs are
difficult to determine \citep[e.g.,][]{2014arXiv1404.2934S}, and the
selection effects associated with the detectability of FRBs are still being
understood
\citep{2013MNRAS.436L...5L,2014ApJ...789L..26P,2014arXiv1407.0400B,2015MNRAS.447.2852K}.
In addition, DMs and fluences are not the only information available: most
FRBs show significant scattering \citep[e.g.][]{2013Sci...341...53T}, which
in principle can provide additional constraints on their distances and
environments \citep{2013ApJ...776..125M,2014arXiv1409.5766K}.  In terms of
foreground modeling, it is now well-established that the NE2001 model is not
a complete description of the Galactic electron distribution, especially at
high latitudes \citep{2008PASA...25..184G}. While any errors in NE2001 do
not have a qualitative effect on the conclusions of our present study,
improved foreground electron distributions will be needed if FRBs are to be
used as precision probes of cosmology and dark energy
\citep[e.g.,][]{2014ApJ...788..189G,2014PhRvD..89j7303Z}.
In addition, the exact distribution of baryons around galactic halos
is still not well understood and depends on details of its implementation
into numerical simulations \citep[e.g.][]{2015arXiv150302084F}, which
could alter the DM signal expected from intervening, low mass galaxies
\citep[e.g.][]{2014ApJ...780L..33M}. Looking to the
future, not only can we expect larger numbers of FRBs, but we now know that
FRBs are polarised \citep{2014arXiv1412.0342P}. This raises the prospect of
detecting Faraday rotation for FRBs, so that we can simultaneously obtain
both DMs and rotation measures (RMs). Such data can provide direct
measurements of the magnetisation of the IGM \citep{2014ApJ...797...71Z,2015arXiv150107535M}, and hence can
potentially discriminate
between different mechanisms for the origin of cosmic magnetism
\citep[e.g.,][]{ddlm09}.


\section*{acknowledgments}
K.D. and A.M.B. acknowledges the support by the DFG Cluster of Excellence ``Origin and Structure of the Universe'' and the DFG Research Unit 1254 ``Magnetisation of Interstellar and Intergalactic Media''. 
B.M.G. is supported by the Australian Research Council
Centre of Excellence for All-sky Astrophysics (CAASTRO), through project
number CE110001020.
This work was conceived and initiated during the workshop ``Tracing the
Cosmic Web'', held at the Lorentz Center at Leiden University in
February~2014.
We are especially grateful for the support by M.~Patkova through the Computational Center for Particle and Astrophysics
(C$^2$PAP). Computations have been performed at the at the
`Leibniz-Rechenzentrum' with CPU time assigned to the Project `pr86re' as
well as at the `Rechenzentrum der Max-Planck- Gesellschaft' at the
`Max-Planck-Institut f\"ur Plasmaphysik' with CPU time assigned to the
`Max-Planck-Institut f\"ur Astrophysik'.  Information on the {\it Magneticum
Pathfinder} project is available at http://www.magneticum.org.

\bibliographystyle{apj}
\bibliography{master,master3,Literaturdatenbank}

\end{document}